\documentclass[11pt,paper=letter]{scrartcl}

\usepackage{graphicx}
\usepackage{amsmath}
\usepackage{amsfonts}
\usepackage{amssymb}
\usepackage{nicefrac,mathtools}
\usepackage{bm}
\usepackage{amsthm}

\usepackage[numbers,sort]{natbib}
\usepackage{xspace}
\usepackage{xcolor}
\usepackage{enumerate}
\usepackage{verbatim}
\usepackage{subfigure}
\usepackage{bbm}
\usepackage{xr}
\usepackage{threeparttable}

\usepackage[colorlinks,linkcolor=blue,citecolor=red]{hyperref}
\usepackage{hypernat}
\usepackage[bf,format=plain,singlelinecheck=false]{caption}

\usepackage{algorithm}
\usepackage{algorithmic}

\usepackage[justification=centering]{caption}

\newtheorem{assumption}{Assumption}

\usepackage{times}
\def\P{\mathbb{P}}
\newcommand{\iid}{\stackrel{\mathrm{iid}}{\sim}}
\def\E{\mathbb{E}}
\def\Unif{\textnormal{Unif}}

\newcommand\independent{\protect\mathpalette{\protect\independenT}{\perp}}
\def\independenT#1#2{\mathrel{\rlap{$#1#2$}\mkern2mu{#1#2}}}
\def\obs{\textnormal{obs}}

\newcommand\hz[1]{\textcolor{black}{#1}}

\title{ Heterogeneous Treatment Effect Estimation through Deep Learning }

\author{ Ran Chen\thanks{ Department of Statistics, The Wharton School, University of Pennsylvania } 
\and 
Hanzhong Liu\thanks{Center for Statistical Science and Department of Industrial Engineering, Tsinghua University, Beijing, 100084, China} }

\begin{document}
	
\maketitle

\begin{abstract}
 \noindent	\textbf{Abstract}

  \noindent	 Estimating heterogeneous treatment effect is an important task in causal inference with wide application fields. It has also attracted increasing attention from machine learning community in recent years. In this work, we reinterpret the heterogeneous treatment effect estimation and propose ways to borrow strength from neural networks. We analyze the strengths and drawbacks of integrating neural networks into heterogeneous treatment effect estimation and clarify the aspects that need to be taken into consideration when designing a specific network. We proposed a specific network under our guidelines. In simulations, we show that our network performs better when the structure of data is complex, and reach a draw under the cases where other methods could be proved to be optimal.
\ \\

 \noindent \textbf{Keywords}

 \noindent  \hz{Heterogeneous Treatment Effect}, Causal Inference, Machine Learning, Deep Learning, \hz{Neural Networks}

\end{abstract}

\section{Introduction\label{intro}}

In many application domains, including economic, marketing, public policy and personalized medicine, and science, estimating heterogeneous treatment effect is a central task. They want to know what is the specific treatment effect for a given individual when some other information (covariates) of the individual is provided, which could be gender, age, genetic information and CT scan images in health care setting, and customer's profile, age, gender and purchasing behaviors in marketing setting. For example, some pharmaceutical companies cares not only average treatment effect of a new drug, but also the treatment effect of a certain group of people characterized by their covariates, so that they can have a better knowledge of whom are the people that should take the drug.


Recently, with growing accessibility of data of various form, estimating heterogeneous treatment effect becomes more and more important and achievable. Researchers have proposed various ways of using machine learning tools to estimate heterogeneous treatment effect, including Lasso \cite{tian2014lasso,jeng2018debiaselasso}, support vector machines \cite{imai2013svm}, causal tree \cite{athey2015machine}, causal forests \cite{wager2017estimation}, ensemble methods \cite{grimmer2017ensemble}, meta learners \cite{hill2011bart,green2012bart,kunzel2017meta}, and deep learning \cite{shalit2016,kunzel2018transfer}, seeking to provide more accurate heterogeneous treatment effect estimate under certain assumptions for different causal problems, among which, linear regression based method and tree based methods are more recognized and form the two streamlines. (see below for a more detailed review)



\hz{In this paper, we propose a new framework for estimating heterogeneous treatment effect using deep learning. We reinterpret heterogeneous treatment effect estimation and analyze both the advantages and issues taken into account when integrating neural network into heterogeneous treatment effect estimation. We develop a specific causal network configuration for this purpose when informative image data are available in randomized experiments or in observational studies. The proposed method is very simple and is convolutional neural network (CNN) based, but has already shown a huge advantage in heterogeneous treatment effect estimation with complex data.}


\subsection{Definitions and Problem Setting}
\hz{We consider a setup where there are $n$ individuals in the experiment (or observational study). For each individual $i$, let $T_i$ be the binary indicator for treatment assignment with $T_i=1$ for treated individual and $T_i=0$ for control. Let $X_i\in \mathbb{R}^p$ be a $p$-dimensional vector of baseline covariates.} Following the Neymann-Rubin potential outcome framework \cite{neyman1923potential, rubin1974estimating}, 
\hz{there exist  two potential outcomes for individual $i$, $Y_i(1)$ and $Y_i(0)$, representing the response when $i$ is exposed to treatment and control respectively.  The individual level treatment effect is defined as the difference in potential outcomes,
$$
\tau_i = Y_i(1)-Y_i(0).
$$
In the experiments (or observational study), we cannot observed $Y_i(1)$ and $Y_i(0)$ simultaneously. The realized and observed outcome for individual $i$ is a function of treatment assignment and potential outcomes:
$$
Y_i^\obs = T_i Y_i(1) + (1 -  T_i) Y_i(0).
$$
}
\hz{Our observed data consist of the triple $(Y_i^\obs,X_i,T_i)$, for $i=1,\cdots,n$, which are assumed to be an independent and identically distributed (i.i.d.) sample drawn from a super probability distribution, that is $(Y_i^\obs,X_i,T_i)\iid \P $.}

Our goal is to estimate Individual level Treatment Effect (ITE), $\tau_i = Y_i(1)-Y_i(0)$, as accurately as we can when we are given a new item with $X_i$ observed. But this is not possible without extremely strong assumptions \cite{kunzel2017meta} . A naive choice is to estimate the Average Treatment Effect (ATE), $\tau=\E(Y_i(1)-Y_i(0))$, as a substitute for $\tau_i$. However, this approach does \hz{not} consider the individual properties given in $X_i$. Another way people often use is to estimate Conditional Average Treatment Effect (CATE) \hz{defined by
$$
\tau(x) = E[Y_i(1)-Y_i(0)|X_i=x],
$$ 
}as a substitute for $\tau_i = Y_i(1)-Y_i(0)$, using information in $X_i$. \hz{The CATE is often regarded as heterogeneous treatment effect which is the} estimand \hz{ of this work.} 
 
\hz{To make CATE identifiable, researchers often assume ``unconfoundedness" \cite{rubin1983unconfoundedness} and ``population overlap" \cite{khan2010overlap}.
\begin{assumption}[Unconfoundedness]
$$
T_i \independent \left(  Y_i(1),Y_i(0) \right) | X_i .
$$
\end{assumption}
\begin{assumption}[Population Overlap] 
$$
P(T_i=1|X_i=x) \in (0,1) \quad \textnormal{with probability 1.}
$$
\end{assumption}
}

However, it should be noticed that $X_i$ is what we observed, so that it does not necessarily carry enough pertinent information about $\tau_i$ due to two facts: (1) We cannot assert we can observe all the measures related to $\tau_i$; and (2) $X_i$ can be very noisy and noise does not tell us much about the $\tau_i$. Therefore, the distribution of $(\tau_i | X_i=x)$ could be heavy tail due to the noise, or only have a small concentration around the likely $\tau_i$ due to missing information. In these cases, which are very likely, CATE is not a good substitute for Individual Treatment Effect, conditional mode may be a better choice. We will elaborate on this more later in a specific case, but for now, what's under consideration is CATE.

\subsection{Related Works}
Many endeavors have been made to borrow strength of machine learning methods \hz{for estimating different kinds of treatment effects}. Two main streamlines are linear regression based methods and tree based methods. 

\hz{Linear regression or penalized linear regression have been adopted in estimating average treatment effect (ATE) with the goal of either improving the estimation accuracy in randomized experiments \cite{freedman2008regression,lin2013,bloniarz2015lasso} or removing the bias caused by confounders in observational studies \cite{belloni2013program}. As critiqued in \cite{freedman2008regression},  adjusted regression, which regresses the observed outcome $Y_i^\obs$ on the treatment indicator $T_i$ and covariates $X_i$, can harm the efficiency of ATE estimation even in completely randomized experiments. Freedman's critique can be resolved by adjusted regression with interaction which incorporates interactions between the treatment indicator and covariates into the linear regression \cite{lin2013}. Researchers also proposed using penalized linear regression for high dimensional setting to estimate heterogeneous treatment effect. \cite{tian2014lasso} advocated solving the Lasso on a reduced set of modified covariates for detecting treatment effect variation across covariates, but they assume that the treatment is randomly assigned and the observed outcomes follow a linear regression model with modified covariates and treatment-by-covariates (modified) interactions. \cite{jeng2018debiaselasso} proposed an asymptotically unbiased estimator for personalized treatment decision based on Lasso solution for the coefficients of interactions between treatment indicator and covariates. The proposed method is valid regardless of the baseline function of the regression model is misspecified or not, but what they really infer is the linear projection of the CATE on the covariates, not the CATE itself.}


\hz{Next we review the tree-based methods. The causal tree introduced in \cite{athey2015machine} integrates the notion of treatment effect into the tree building and fitting procedure and is designed to estimate CATE. \cite{wager2017estimation} proposed causal forest which ensembles causal trees to both estimate CATE and construct confidence interval for it. \cite{kunzel2017meta} proposed Meta-learners, a set of methods integrating treatment indicator into machine learning methods, mostly tree based methods. Meta-learners include S-learner, T-learner and X-learner. The S-learner is to estimate the observed outcome using all of the covariates and the treatment indicator. We denote the estimate as $\hat \mu (x,T)$. The CATE can be estimated as the difference of $\hat \mu (x,T)$ by changing the treatment indicator from 0 to 1, with the covariates fixed, i.e., $\hat \mu(x,1)  -  \hat \mu (x,0)$. Regression tree \cite{athey2015machine} and Bayesian Additive Regression Trees (BART) \cite{hill2011bart,green2012bart} have been studied as base learners of S-learner. The T-learner commonly takes two steps for estimating CATE. First, it uses base learners such as regression trees to estimate the conditional expectation of the potential outcomes given the covariates
$$
\mu_1 (x)  = E[ Y_i (1) | X_i = x], \quad \mu_0 (x)  = E[ Y_i (0) | X_i = x].
$$
Second, it takes the differences of these two estimates \cite{athey2015machine}. The X-learner builds on the T-learner and uses the imputed individual-level treatment effect as response variable to obtain two CATE estimates from the treatment and control group respective. The final CATE estimate is a weighted average of the two estimates \cite{kunzel2017meta}. The X-learner is particularly efficient when the number of individuals in one group (often the control group) is much larger than the other.}


\hz{Aforementioned work have also validated their methods theoretically under specific settings, either assume the observed outcomes follow a regression model with treatment-by-covariate interactions or assume the potential outcomes $Y_i(1)$ and $Y_i(0)$ follow two separate regression models and the underlying mean regression functions enjoy Lipschitz conditions or certain degree of smoothness.}

However, with the growing accessibility of data of various form, the structure of data could be complex. In target marketing, data includes user's profile photo, posts, demography information and purchase behaviors. In a medical setting, observable covariates are not restricted to blood pressure, hemogram, gender, image data like CT scan can also be collected. Therefore, data can be of very high dimension, can have complex structure, and can be very noisy. The simple models may fail to be representative in a lot of applications as mentioned above. 

\hz{Endeavors have recently been made to borrow strength of deep learning methods for estimating treatment effect. \cite{pham2017deepcausal} used a deep causal inference approach to measure the average effects of forming group loans in online non-profit Microfinance platform. The adopted deep neural network was borrowed directly from natural language processing and this paper did not tackle the problem of estimating CATE. \cite{shalit2016} proposed estimating CATE through representation learning --- learning a balanced representation before continue to the rest of the algorithm. They gave a generalization bound, and provided a specific realization of algorithm using a neural network. However, the generalization bound only bounded the mean squared error (MSE) of the CATE estimation by the representation unbalancedness, the unexplainable variance of potential outcomes and the sum of MSE of CATE estimation on treated group and control group, which did not give much information. Although good representation is needed for complex data, a balanced representation does not necessarily lead to a better CATE estimation. \cite{kunzel2018transfer} developed new algorithms using transfer learning with neural networks for CATE estimation, but their main goal was to deal with different data sources that are related to the same underlying causal mechanisms.}



In this paper, we propose a framework of integrating neural network into heterogeneous treatment effect estimation in general, specify the gains and issues needs consideration when using this approach in general, and we give a specific CNN based neural network configuration to better estimate CATE. It works in both randomized experiment and observational study. It uses the diverter mechanism we developed as a way to incorporate the treatment assignment indicator for predicting the outcome. The diverter mechanism automatically enables both information share and separation in treatment and control group, and in our diverter mechanism, treatment indicator is not by nature restricted to 0 and 1, it can take continuous value, like dose. Our causal network shows huge advantage in simulated image data where the topological structure is related to heterogeneous treatment effect. Moreover, our causal network can easily combine the strength of other time-proved network structures (e.g., Recurrent Neural Network(RNN)) and other non neural neural network based structures (due to the expressiveness of neural network) to fully borrow the strength of neural networks into heterogeneous treatment effect estimation with various form of data.



\subsection{Reinterpretation of the problem}

The problem of heterogeneous treatment effect estimation is defined on some model assumptions that captures the essentials of the real world problem, but with concessions and approximations. All the aforementioned methods are by nature a designation of a computation procedure, resulting in an output, no matter what model assumptions are. Therefore, the task becomes, coming up with a computation procedure that can result in an output ``closer", as measured by \hz{mean squared error (MSE)} or other criterion, to ``truth" or its proxy, which is CATE in our problem setting, under the model assumptions we defined above. However, the model assumptions of the problem is too general for this goal to be fully achieved, a reasonable goal is to achieve good performances in a reasonable wide arrange of settings, among which, regression models with Lipschitz conditions are only a small part. Some preferable characteristics of such methods are:
\begin{itemize}
\item Can fit different models well without knowing the model specification,
\item Can achieve consistency when sample size \hz{grows} to infinity,
\item Computationally \hz{feasible}.
\end{itemize}

The second property is hard to achieve theoretically when the first is achieved empirically, as the setting could be complex enough to resist statistical and mathematical analysis. The third is also in conflict with the second one to some extent, when one goes for optimal convergence rate in some problems \cite{cai2017}. Therefore, in this work, we do not put any of the criterion as dominant, all the three are important.

\section{Causal Networks}

\subsection{Integrating Neural Network into Heterogeneous Treatment Effect Estimation}

Neural network has proved to be successful in image tasks and text tasks, due to its amazing expressiveness and the ability to deal with structured data, which motivates borrowing its strength into heterogeneous treatment effect estimation.

A working neural network has two basic elements: network structure and training paradigm. 

The neural network structure \hz{is} normally composed of input layer, several hidden layers and an output layer. Each Layer is composed of operators: linear transformation or piecewise linear transformation (max pooling), activation function, and batch normalizing (stabilize the input data in each layer). One can have an output with a neural network when parameters are fixed. With a given pre-specified loss function, the aim is to optimize the loss function, \hz{where} the training paradigm comes into consideration. Common training paradigms are back propagation based training paradigm, Adam and \hz{Stochastic Gradient Descent (SGD)} are popular ones.

\subsubsection{Expressiveness of Neural Networks} 
Neural networks can express a wide range of relationships through a specific realization of parameters. For example, only a linear transformation operator is enough to express linear relationship, as could be expressed in linear regression based methods. When there are more layers with activation functions, whether linear or not, linear relationship can also be approximated as long as the activation function is differentiable on an interval, mean while, nonlinearity could also be detected as long as the activation function is second-order differentiable. Our exploratory simulation shows that two layer neural network with nonlinear activation function can detect nonlinearity (see appendix).

Though neural network output is a continuous function of input, the step function could be approximated by two Relu (a kind of activation function) with opposite direction. Forest structure with a given number of trees can also be approximated by adding an averaging layer (which is a linear transformation) in the end.

The expressiveness of neural network makes it possible to incorporate several methods in one network structure, the conventional model selection procedure is some what incorporated in the training procedure w.r.t. minimizing the loss function. The notion of model selection in neural network setting becomes selecting a neural network structure.

\subsubsection{Computational Burden}
A key issue in exploiting neural network based method is it's computational burden --- it does not compute the exact solution and the computation time could be long.

Computation time primarily depends on the number of nodes and complexity of the function used in that node, so moderately large neural networks do not require much time for one updating procedure and it is linear to sample size. On the other hand, for random forest type methods, the computation time depends both on number of trees in the forest and the tree fitting procedure, both of which commonly grow with sample size.  Besides, with the development of neural network in computer science, chips for computing neural networks are under development, in which situation the computation of basic operators becomes a single instruction like ``==" and ``!=". The problem of computational time is not an issue then.

What can not be offset by the recent development in computer hardware is that the problem is not exact solution. In our exploratory simulation, we find that with increasing iteration and access to new data, this problem can be allieviated. The problem exists, but to mild extent.

\subsubsection{End to End Approach}
Another characteristic coming along is that this is an end to end approach, this gives the method a huge space for accommodating to different data form. Data of text form, image form, traditional covariates could be considered together in one neural network. Though images data normally use CNN based neural network, text data normally use RNN based network, two kinds of network structure could be connected and merged into a big one and which is not the case for other methods.

\subsubsection{Utilizing Treatment Group/ Control Group Information}
A key task in integrating neural network in heterogeneous treatment effect estimation is figuering out how to use the information in treatment group and control group.

Treatment group and control group shares a part of information, as the data of the two are of the same form and the outcome depends not only on what treatment is given but also on which individual it is. Two groups also has separate information due to the different treatment, which is what we want to get. How to achieve this information sharing and information separation is a key question.
 
\textbf{Diverter.} Here we design a ``diverter", which ``diverts" the information flow of two groups (though our way of integrating treatment information does not restrict treatment indicator to be binary, it could be continuous):

\begin{itemize}
\item Compute according to first several layers of network with covariate $X_i$ to be the input, getting $\mathbf{f}(X_i)$,
\item Control Flow diverter:\\
	    Diverting mechanism: $\mathbf{f_c}(X_i)  = \max(1- \textnormal{sigmoid}( \max(0, \mathbf{f}(X_i)+T_i) ), 0)   $ \\
		Keep going after diverting : $\mathbf{\tilde{f_c}}(X_i) = \mathbf{g_c} \circ \mathbf{f_c}(X_i) $ 
\item Treatment Flow diverter:\\
		Diverting mechanism: $\mathbf{f_t}(X_i) = \textnormal{sigmoid}( \max( 0, \mathbf{f}(X_i)+T_i-1))  $ \\
		Keep going after diverting: $\mathbf{\tilde{f_t}}(X_i) = \mathbf{g_t} \circ \mathbf{f_t}(X_i)   $
\item Merge two flows:\\
		Adding the two flow up to be the output:\\
		$\hat{y}(X_i) = \mathbf{\tilde{f_c}}(X_i) + \mathbf{\tilde{f_t}}(X_i) $
\end{itemize}

\textbf{Intuition of diverter.} The intermediate output $\mathbf{f}(X_i)$ has reasonable range, adding the treatment assignment indicator separate the range of control group and treatment group to some extent, so it is possible to extract the part with ``mostly control group", and the part with ``mostly treatment group", resulting in control flow diverter and treatment flow diverter. After the flow are diverted, they keep going on their computation or transformation. In the end, they are added up together, two flows merges.

With diverter, both information sharing and information separation are incorporated and the extent of information separation are learned through training, as we do not impose a hard separation.


\subsection{Our Causal Network}

\begin{figure}[!h]
\centering
\includegraphics[width=0.8\textwidth]{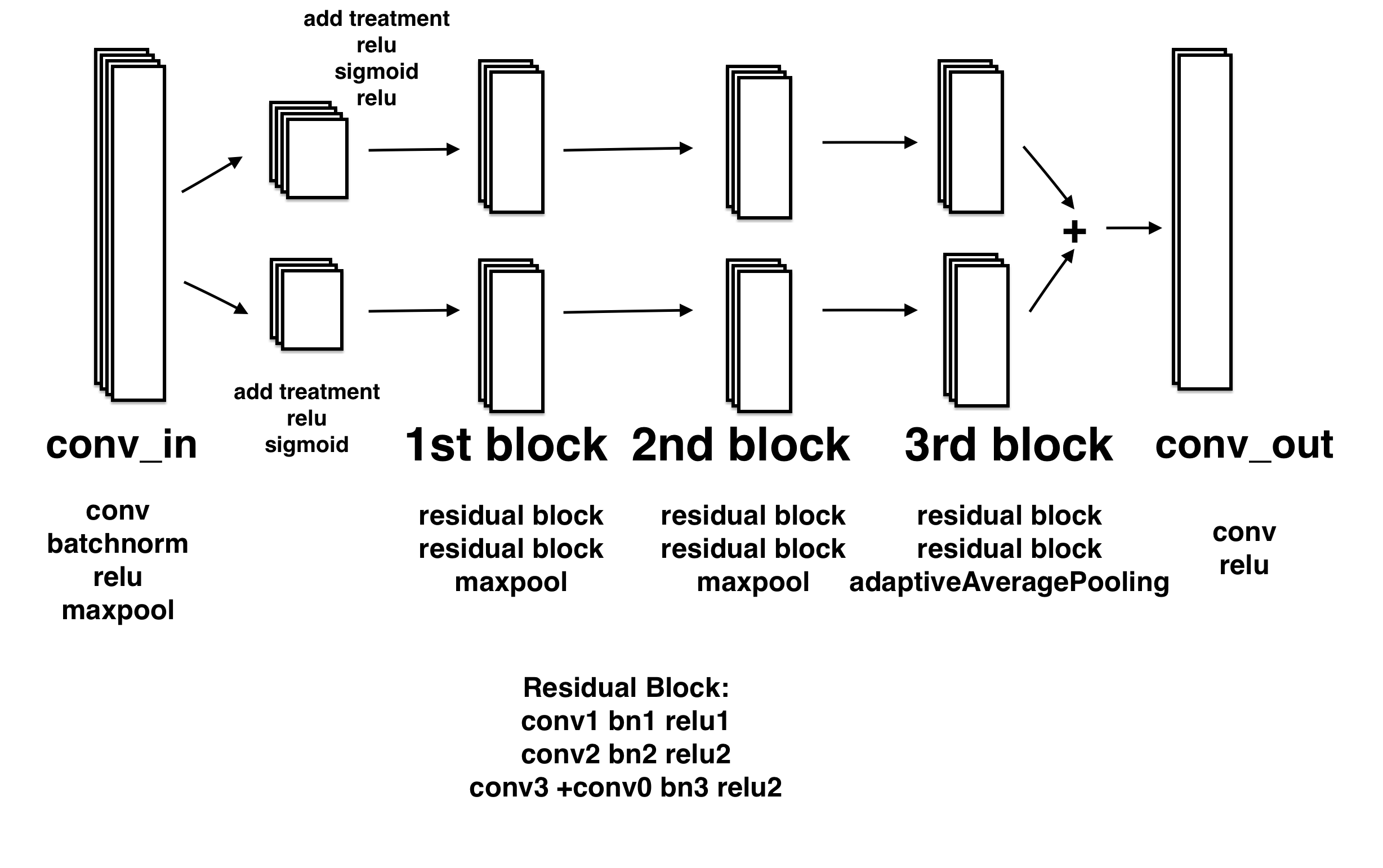}
\caption{Causal Network}
\label{fig:1}
\end{figure}

Under the guidance we discussed above, we designed a simple neural network with a diverter, as shown in Figure \ref{fig:1}. The network is composed of \hz{an} input block, a diverter, three processing block and an output block. This neural network is simpler than the simplest standard neural network, like Alexnet. Parameters in each convolution layer is only the convolution kernel, which is shared across different location of the image tensor.

We use standard SDG for our training procedure. And we use L2 loss function of the outcome to train our network. Our philosophy here is to achieve good performance on CATE estimation through good performance on outcome estimation.


\section{Simulation}

In this section, we test our causal network in various settings and compare it with other methods, namely adjusted regression, adjusted regression with interaction term, \hz{S-learner and T-learner with standard random forest \cite{kunzel2017meta} as base learner.} 

\hz{The first part of this section} is on image data with treatment effect being related to its topological structure, where we try to make the input similar to tumors. We show that our causal network is able to detect the treatment effect information incorporated into the image through topological structures, while all other methods totally fail. We also discussed a little bit on how to define estimand that better quantifies heterogeneous treatment effect in the setting where $X$ is noisy, \hz{for example, $X$ is determined by a hidden variable and additive noise,} which is very likely in reality. 


The second part is devoted to the setting where \hz{the potential outcomes} are generated according to the following schemes: \hz{linear function, polynomial function, given trees, and given neural networks of covariates,} some of which provably favor linear regression based methods or tree based methods. The covariate is still image, which is of high dimension when seen as a vector, and signal to noise \hz{ratio} is less than 10. Therefore, these tasks are very difficult. Our results show that all the methods do not perform well, and their performance are of the same order.

We also choose difference sample sizes for training data to see how the performance vary with sample size. Testing dataset is always the same one with sample size 10000, which is independent to training dataset. \hz{We compare the mean squared error (MSE) $E\left( \hat \tau(X) - \tau(X) \right)^2 $ of different methods, where $\hat \tau(x)$ is an estimate of the CATE $\tau (x)$.}

\subsection{Image data}
\subsubsection{Noiseless Covariate}

\hz{
\textit{Simulation Setting.}\\
Data generating scheme is as follows: for $i=1,\cdots,n$, we independently
\begin{itemize}
\item generate $X_i$ as an 32 * 32 image \hz{containing a circle with random radius $R_i$ and random center at the origin $O_i$},
\begin{itemize}
\item radius $R_i$ is generated from uniform distribution on the interval $[0,16]$, i.e., $R_i \sim \Unif(0,16)$,
\item origin $O_i$ is generated from uniform distribution on the $32 \times 32$ pixel locations, i.e., $O \sim \Unif (0,32)*(0,32) $, and it is independent of $R$, i.e., $ O \independent R$,
\item for pixels inside the circle defined above, pixel value equals $180$,
\item for pixels outside the circle defined above, pixel value equals $0$,
\end{itemize}
\item generate potential outcomes from normal distribution: $Y_i(0) \sim N(0,1)$, $Y_i(1) \sim N(R_i,1)$,
\item generate treatment assignment indicator $T_i$ from binomial distribution, $T_i \sim B(1,0.5)$, which is independent of $(X_i, Y_i(0),Y_i(1))$.
\end{itemize}
}

\begin{figure}[!h]
\centering
{\includegraphics[width=0.5\textwidth]{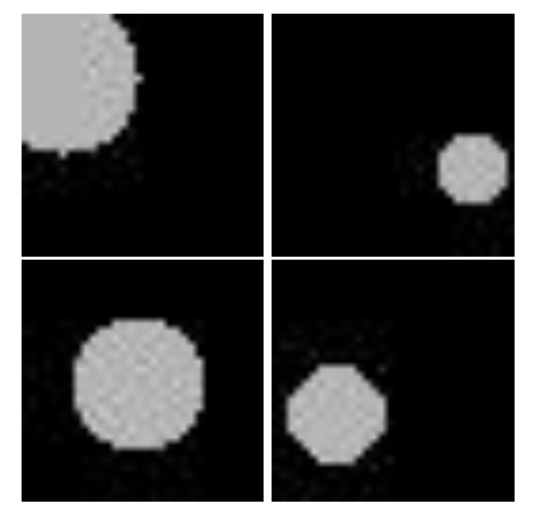}}
\caption{Noiseless Images}
\label{fig:2}
\end{figure}

Figure \ref{fig:2} illustrates how the covariates (images) look like. In this case, $X_i$ contains the pertinent noiseless information about the treatment effect, and $\E(Y_i(1)-Y_i(0)|X_i)=R_i$. We chose training sample size $n$ to be $2000,4000,6000,8000$, and $10000$.


\noindent \textit{Results.} 

\begin{figure}
\centering
\begin{minipage}{.45\textwidth}
{\includegraphics[width=\textwidth]{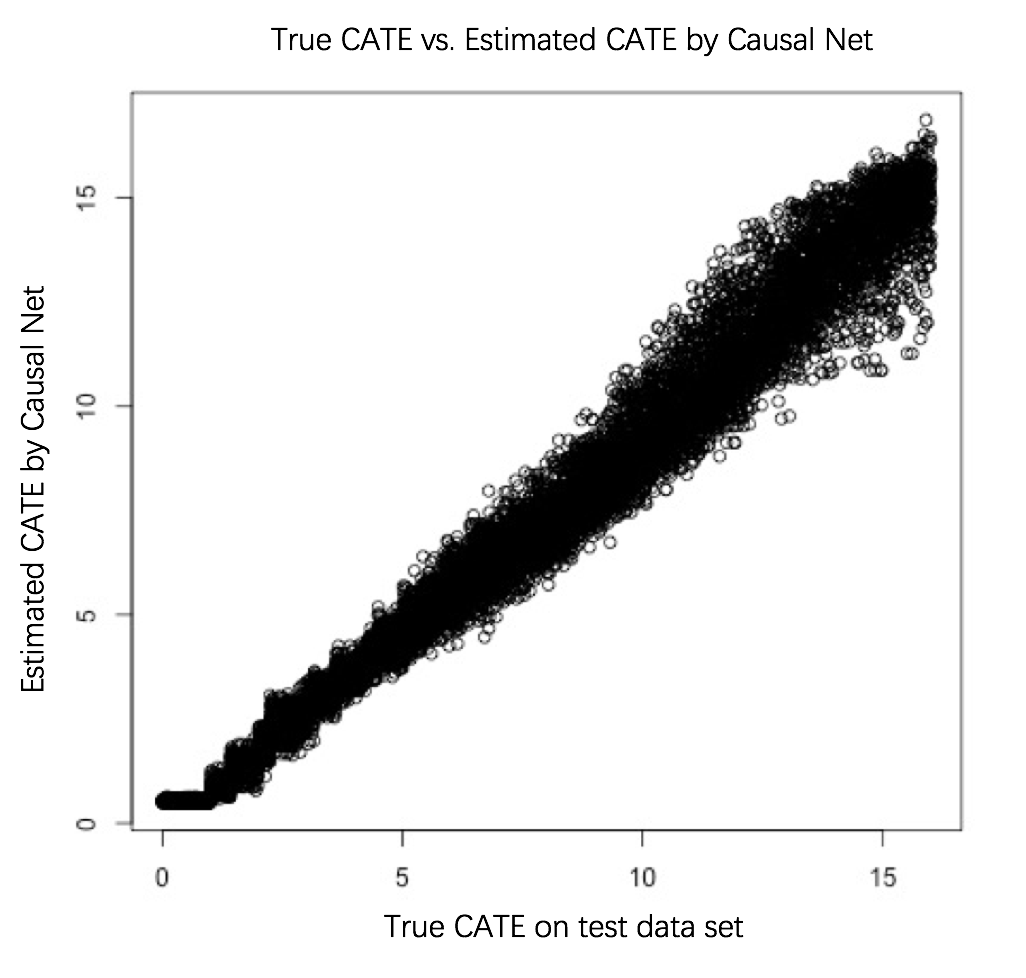}}
\end{minipage}
\begin{minipage}{.45\textwidth}
{\includegraphics[width=\textwidth]{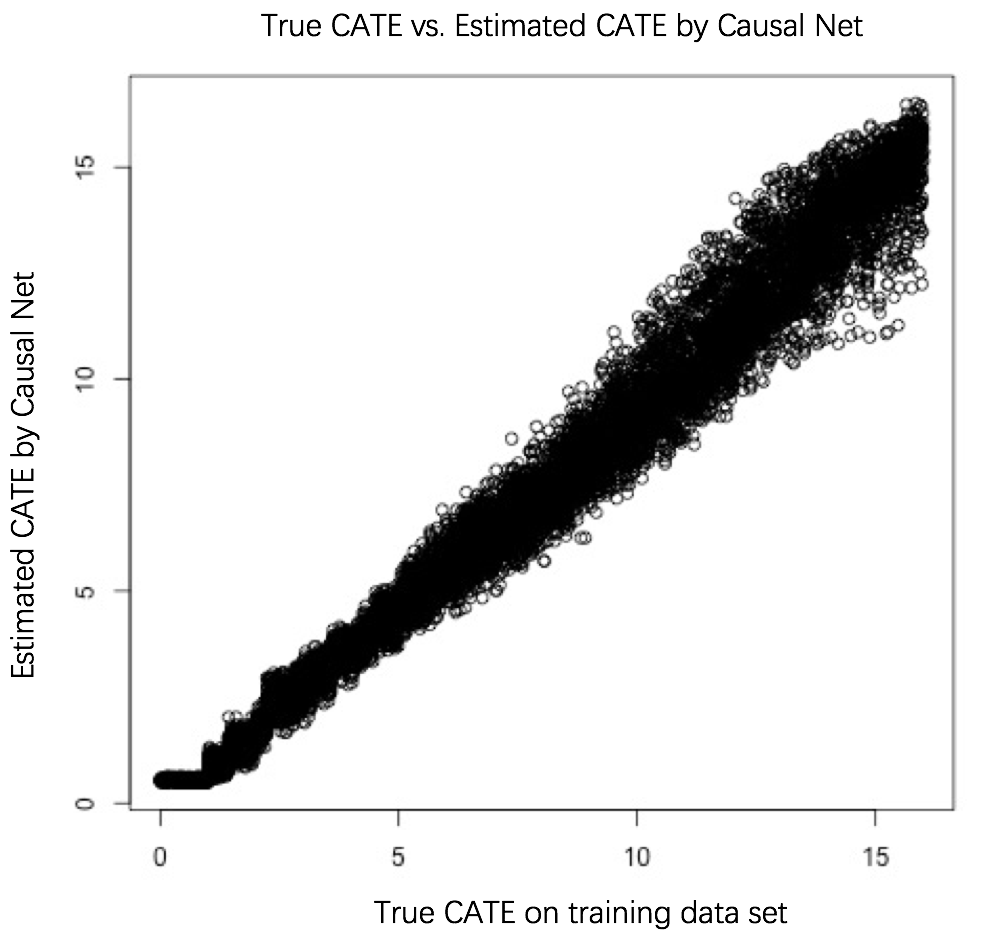}}
\end{minipage}
\caption{CausalNet on noiseless image data with training sample size $n=10000$.}
\label{fig:3}
\end{figure}

\begin{figure}
\centering
\includegraphics[width=.9\textwidth]{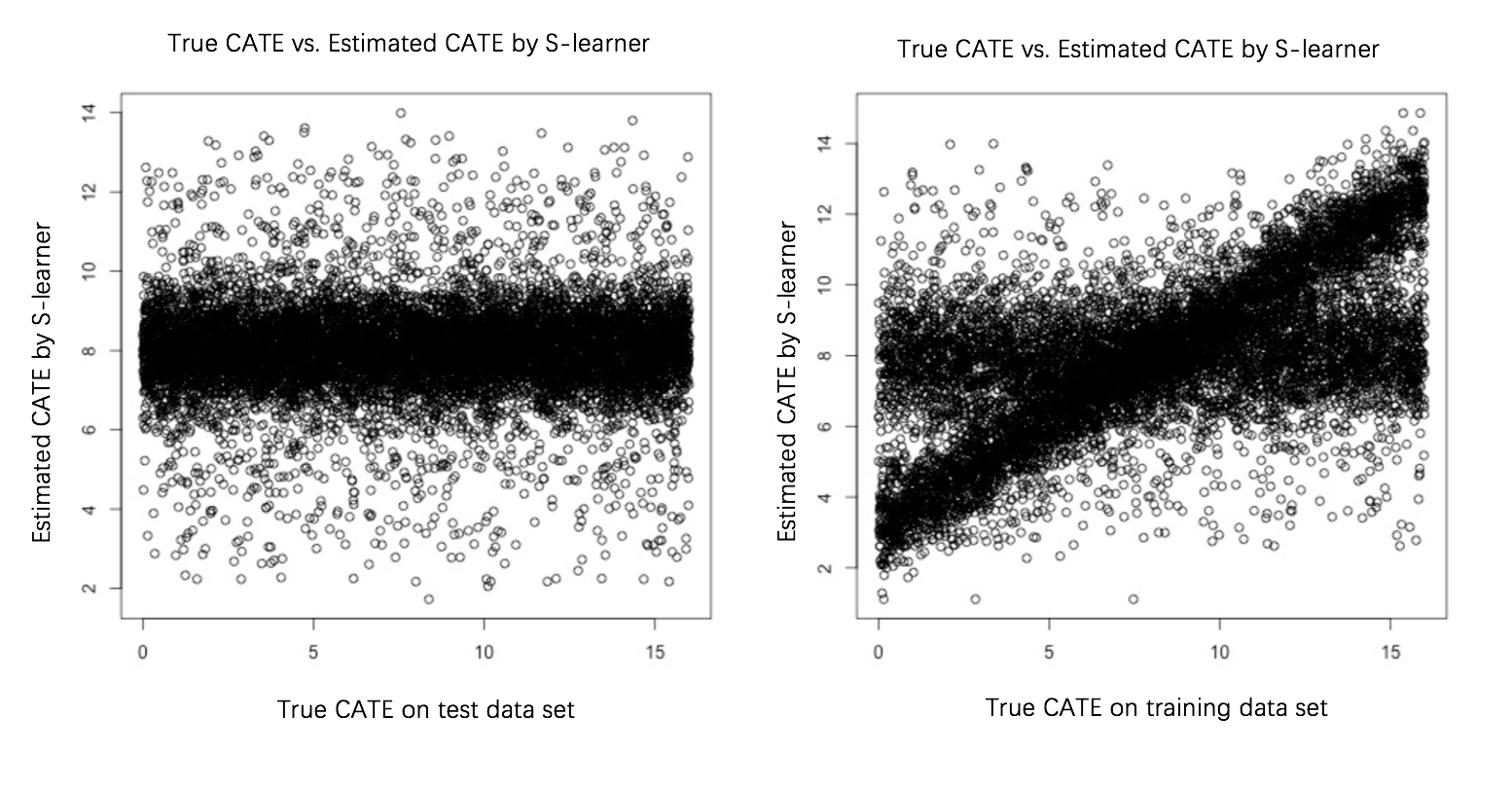}
\caption{S-learner on noiseless image data with training sample size $n=10000$.}
\label{fig:4}
\end{figure}

\begin{figure}
\centering
\includegraphics[width=.9\textwidth]{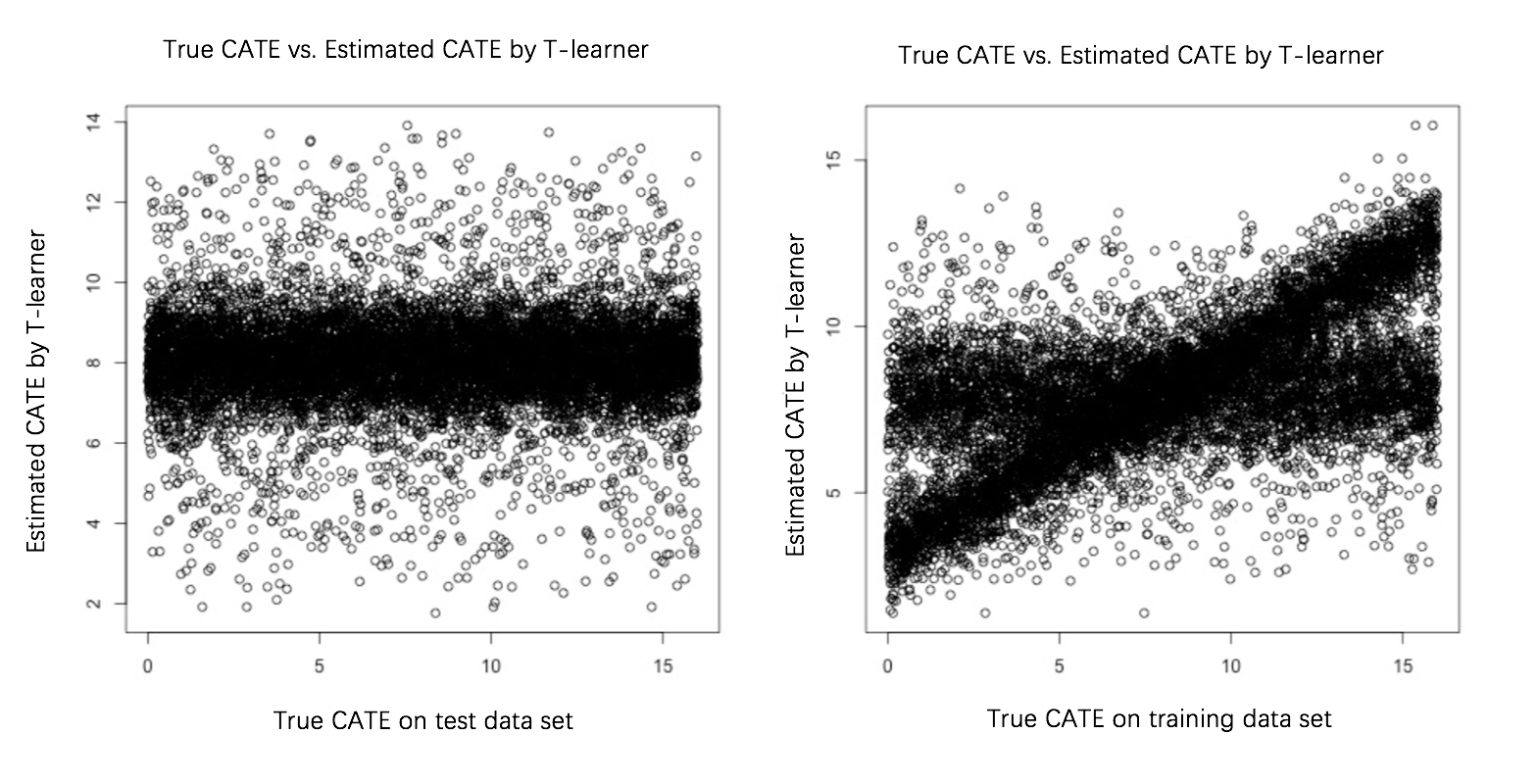}
\caption{T-learner on noiseless image data with training sample size $n=10000$.}
\label{fig:5}
\end{figure}

\begin{figure}
\centering
\includegraphics[width=.9\textwidth]{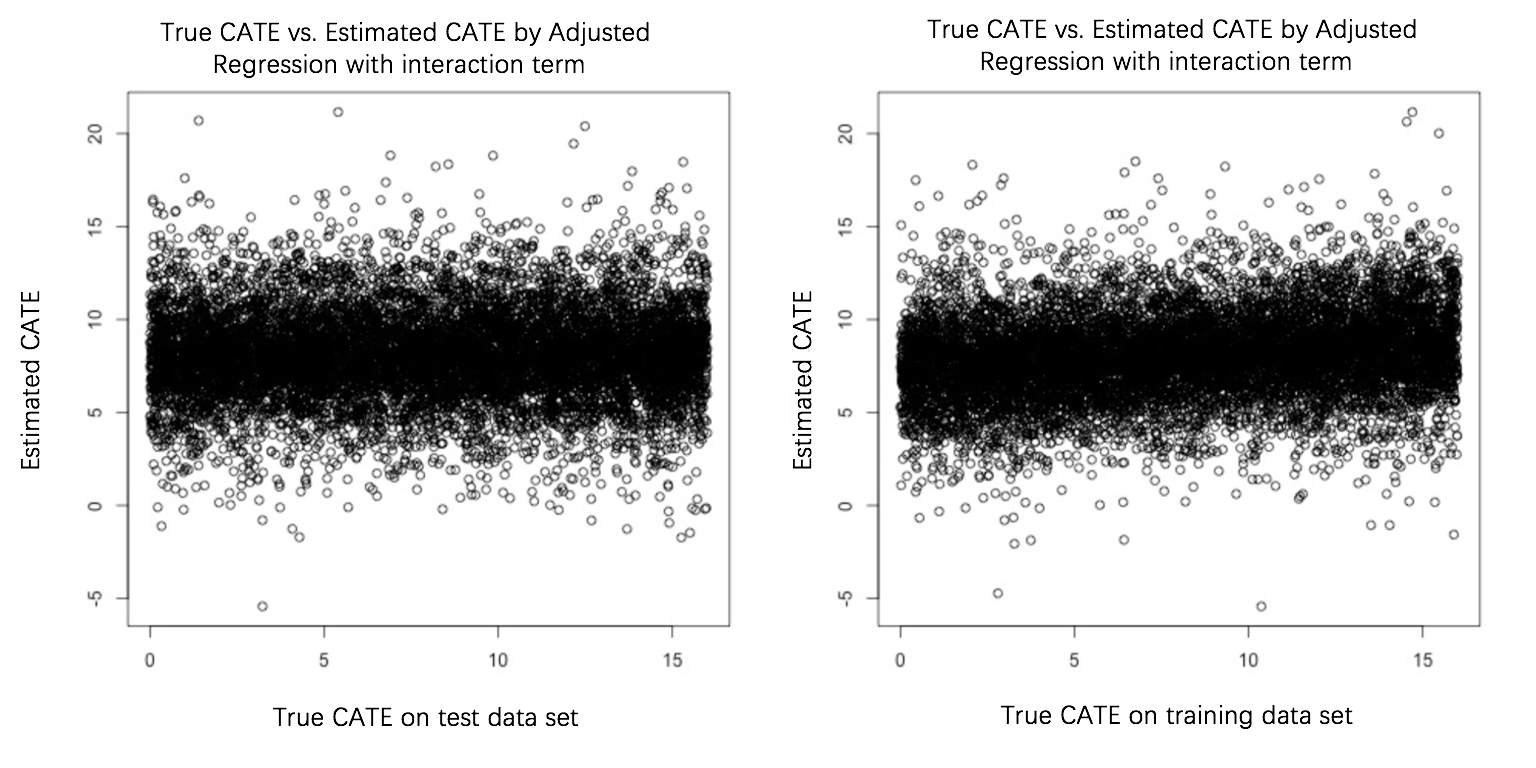}
\caption{Adjusted regression with interaction term on noiseless image data with training sample size $n=10000$.}
\label{fig:6}
\end{figure}

\hz{For training sample size $n = 10000$}, we plot estimation versus true treatment effect on training dataset and test data set. \hz{The results of our causal network (denoted by Causal Net), S-learner, T-learner, adjusted regression with interaction term are shown in Figures \ref{fig:3} - \ref{fig:6} respectively. \hz{In these Figures, the left subplots are results on test data set, while the right subplots are results on training data set.} We can see that causal network performs much better than other methods, where other methods are basically predicting the average treatment effect (around $8$) and cannot detect treatment effect heterogeneity.}

\begin{figure}
\centering
\includegraphics[width=.8\textwidth]{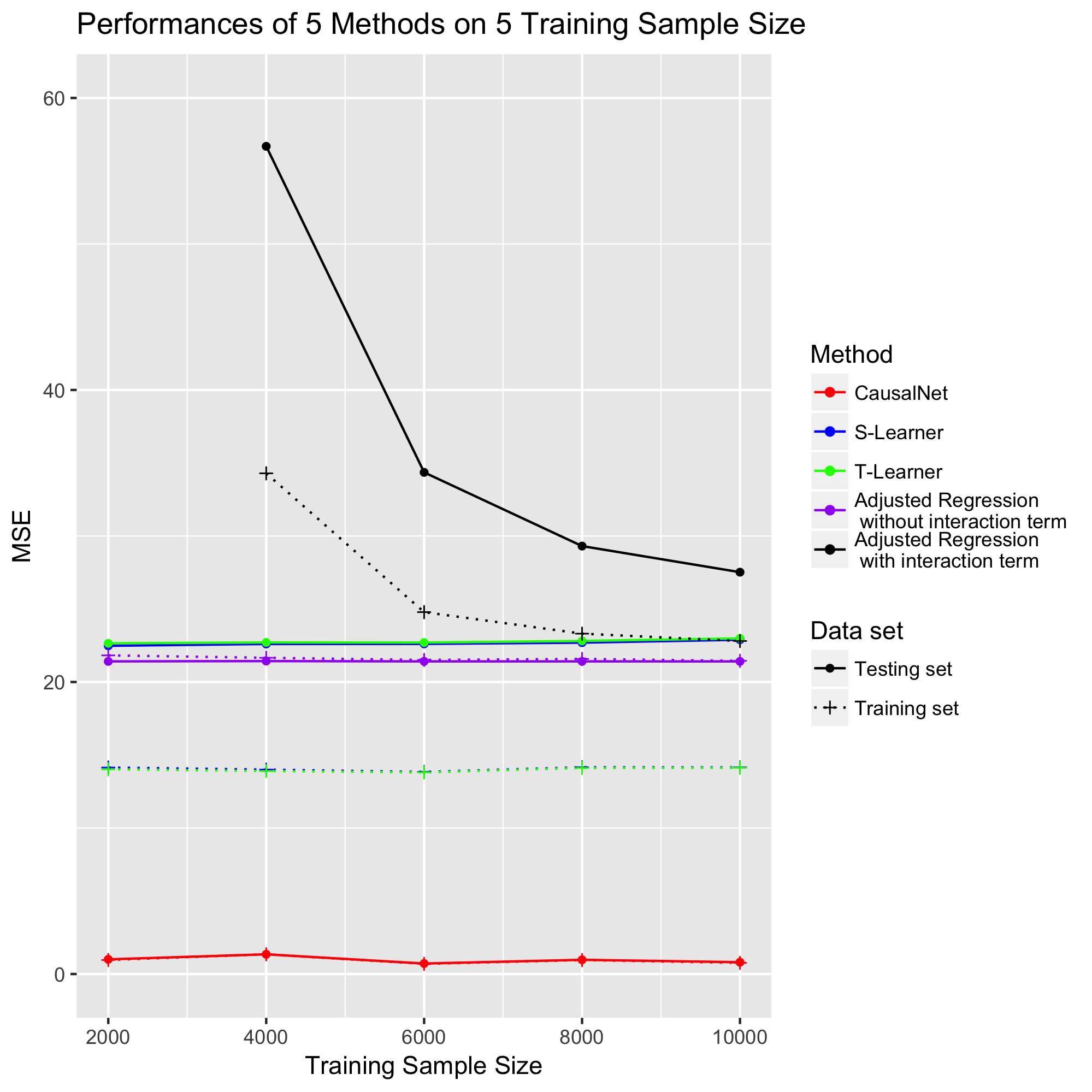}
\caption{MSE of five methods on noiseless image data (on the left) and noisy image data (on the right).}
\label{fig:7}
\end{figure}

When sample size varies, MSE for different methods on both training and testing set are shown in \hz{the left subplot of} Figure \ref{fig:7} (for adjusted regression with interaction, collinearity exists). \hz{We can see that CausalNet improve the estimation accuracy by an order of magnitude for all cases.} Since the true treatment effect is uniform on (0,16), it has variance \hz{$64/3 \approx 21.3$}, and through checking the estimation versus true value plots for different methods on different training sample size, we find that other methods' estimations are almost independent of the true \hz{heterogeneous} treatment effect.

\subsubsection{Noisy Covariate}

Covariates in the previous simulation does not have noise irrelevant to heterogeneous treatment effect. In the real settings however, images are polluted with information that is irrelevant to heterogeneous treatment effect. In this simulation, we add Gaussian noise to see how different methods work. 

However, we notice that $\tau(x) = E(Y_i(1)-Y_i(0) | X_i = x)$ is not necessarily the radius $R$ in this situation, and it changes continuously with the variance of noise. In this situation, the only information that images contain about treatment effect is $R$, so when we are given a new individual, what we really what to get is still $R$. Our substitute for individual treatment effect, CATE, therefore is not a good substitute when the variance of noise is large. Reasons are as follows. Though for $R$ far away from the truth and whatever origin, the probability is \hz{smaller than that of the} true $R$ and origin $O$, the conditional probability of a true $R$ and absurd origin is also small. Since most origins are absurd, advantage of the true $R$ over absurd $R$ is decreased with respect to conditional probability, which is further decreased when conditional expectation is taken. In a very noisy setting, taking the conditional mode to be the substitute may be a better choice. 

\noindent  \textit{Simulation Setting.}

\hz{Detailed data generating scheme is as follows: for $i=1,\cdots,n$, we independently
\begin{itemize}
\item generate $X_i$ as an 32 * 32 image \hz{containing a circle with random radius $R_i$ and random center at the origin $O_i$},
\begin{itemize}
\item radius $R_i$ is generated from uniform distribution on the interval $[0,16]$, i.e., $R_i \sim \Unif(0,16)$,
\item origin $O_i$ is generated from uniform distribution on the $32 \times 32$ pixel locations, i.e., $O \sim \Unif (0,32)*(0,32) $, and it is independent of $R$, i.e., $ O \independent R$,
\item for pixels inside the circle defined above, pixel value is generated from normal distribution $N(180,64)$, and then is truncated to $[0,255]$ and rounded,
\item for pixels outside the circle defined above, pixel value is generated from another normal distribution $N(0,64)$, and then is truncated to $[0,255]$ and rounded,
\end{itemize}
\item generate potential outcomes from normal distribution: $Y_i(0) \sim N(0,0.4)$, $Y_i(1) \sim N(R_i,0.4)$,
\item generate treatment assignment indicator $T_i$ from binomial distribution, $T_i \sim B(1,0.5)$, which is independent of $(X_i, Y_i(0),Y_i(1))$.
\end{itemize}
}

\begin{figure}[!h]
\centering
{\includegraphics[width=0.5\textwidth]{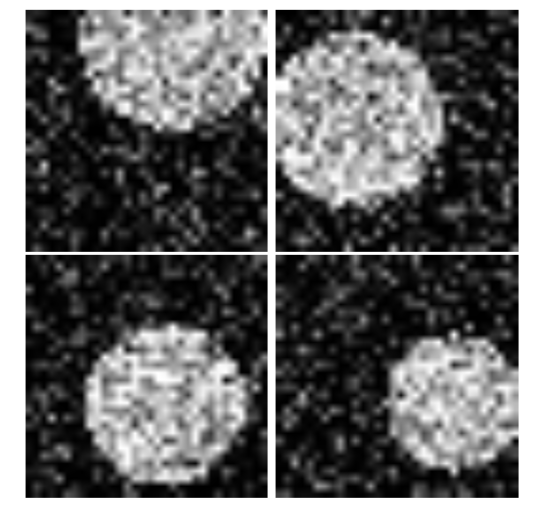}}
\caption{Noisy Images}
\label{fig:8}
\end{figure}

Figure \ref{fig:8} is an illustration of noisy images (covariates). Since the CATE is hard to compute exact in this setting, we take the radius $R$ to be our substitute for heterogeneous treatment effect, and do not change the names in the following figures. 


\noindent \textit{Results.} 

\begin{figure}
\centering
\includegraphics[width=.9\textwidth]{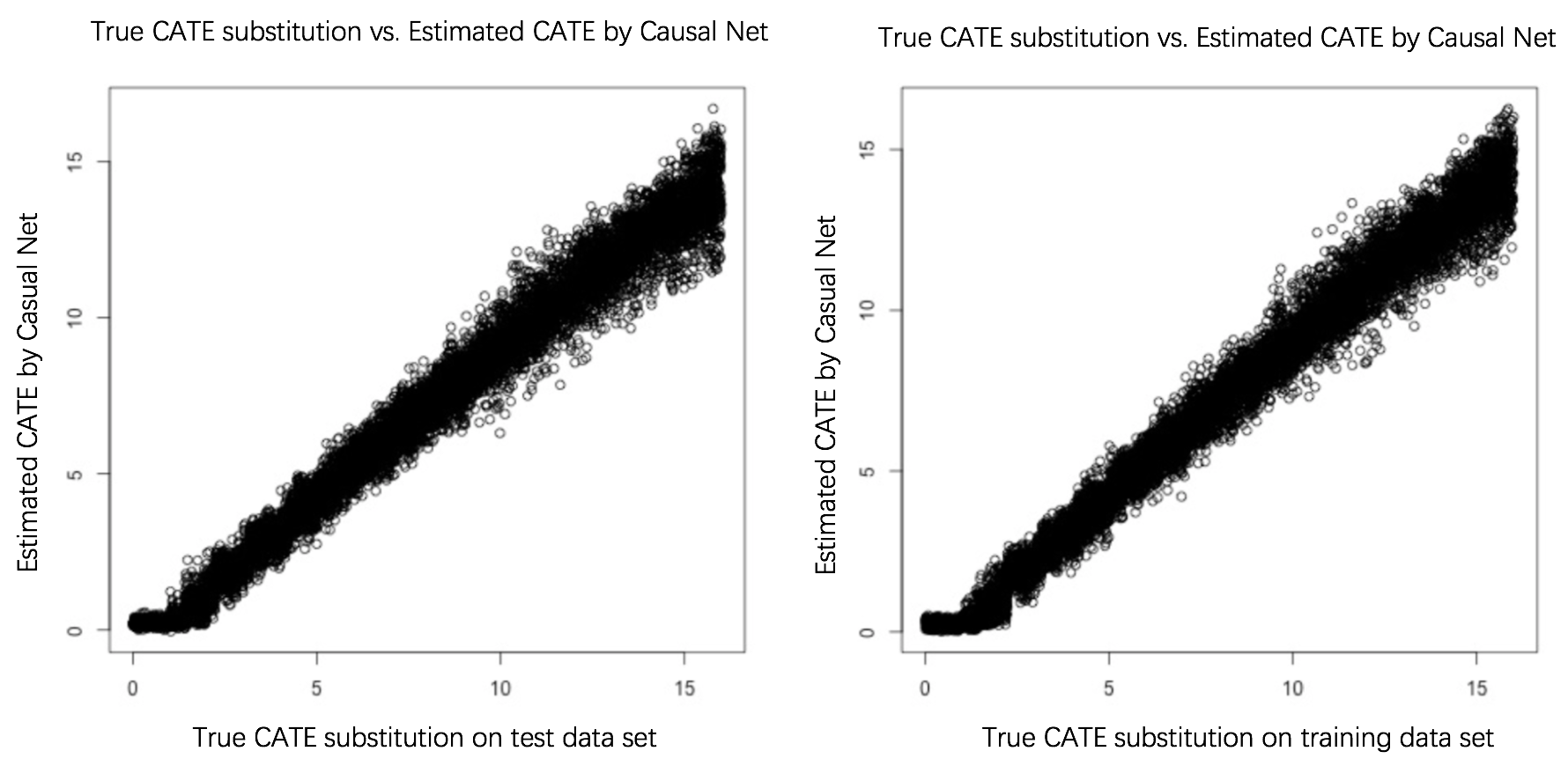}
\caption{Causal Network on noisy image data with CATE substitution and training sample size $n=10000$.}
\label{fig:9}
\end{figure}

\begin{figure}
\centering
\includegraphics[width=.9\textwidth]{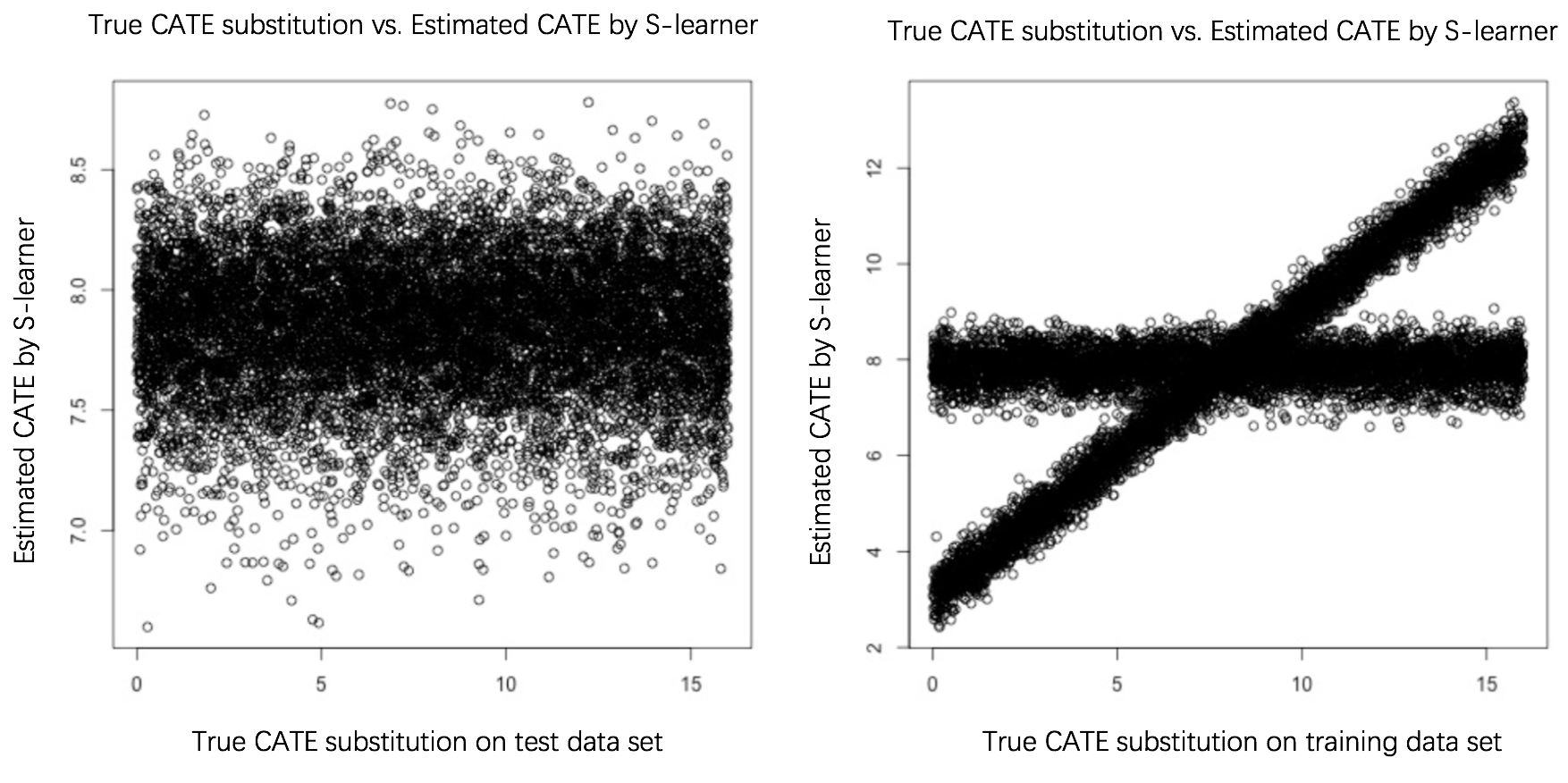}
\caption{S-learner on noisy image data with CATE substitution and training sample size $n=10000$.}
\label{fig:10}
\end{figure}

\begin{figure}
\centering
\includegraphics[width=.9\textwidth]{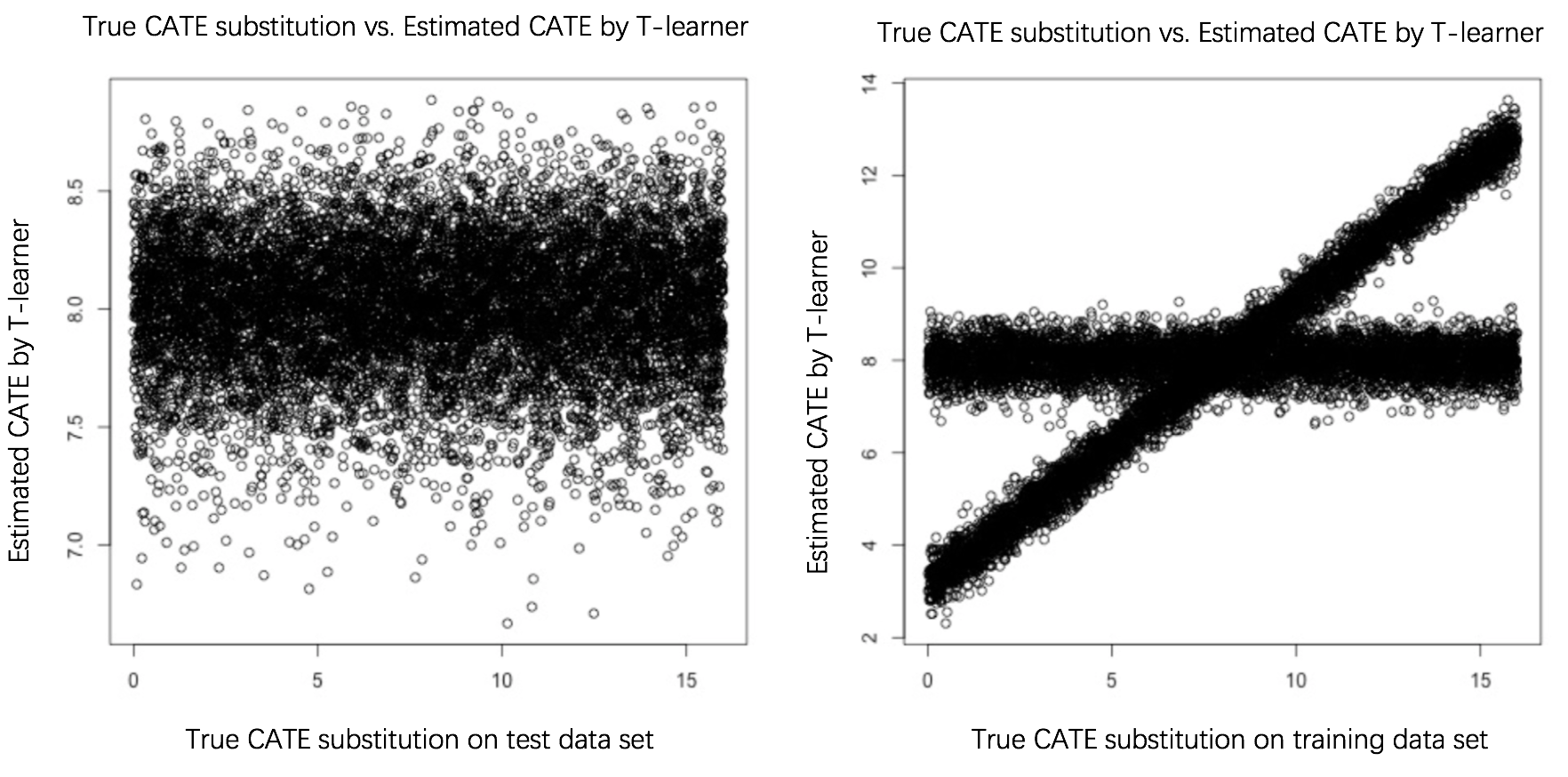}
\caption{T-learner on noisy image data with CATE substitution and training sample size $n = 10000$.}
\label{fig:11}
\end{figure}

\begin{figure}
\centering
\includegraphics[width=.9\textwidth]{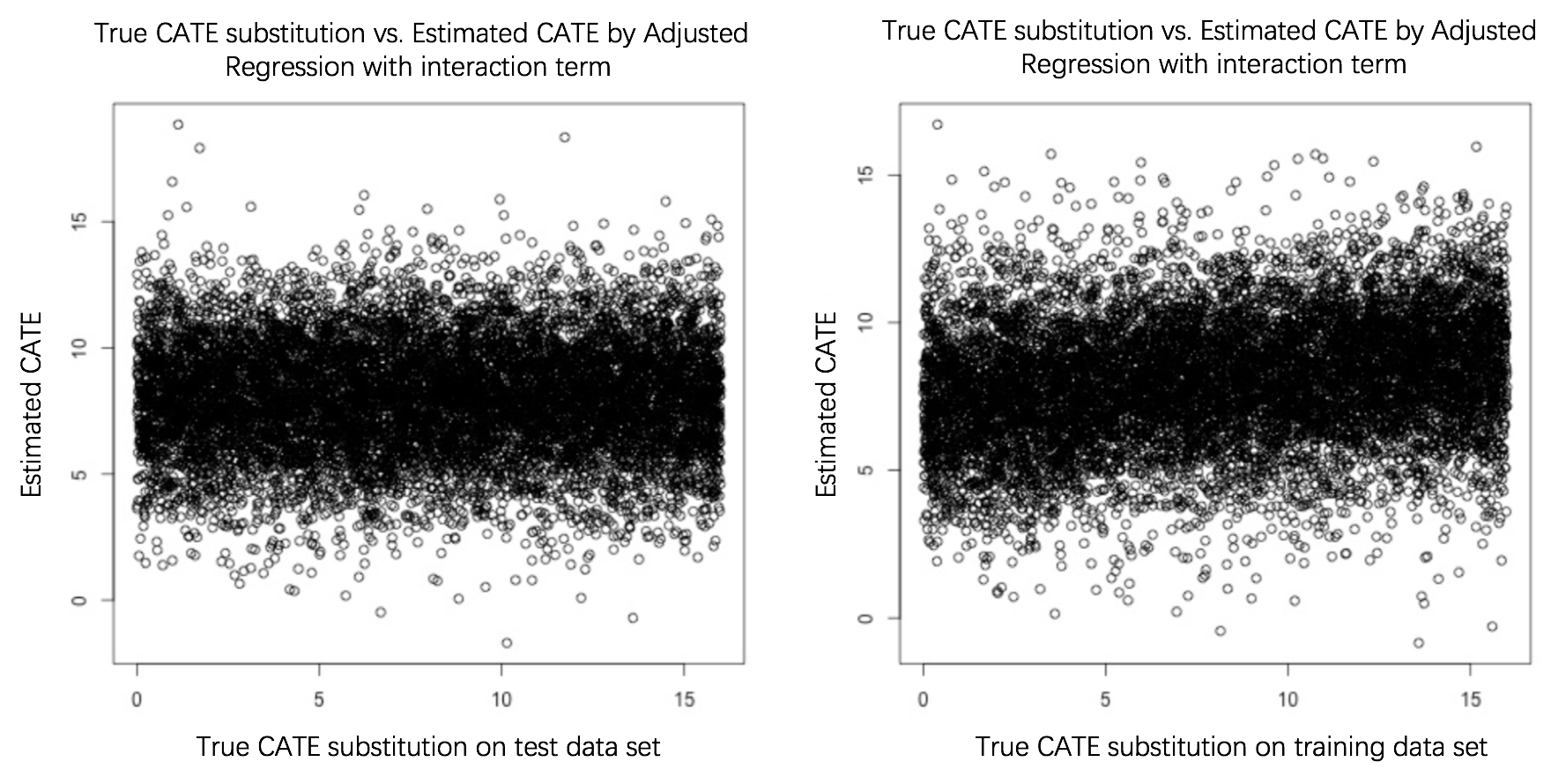}
\caption{Adjusted regression with interaction term on noisy image data with CATE substitution and training sample size $n=10000$.}
\label{fig:12}
\end{figure}

\hz{Figures \ref{fig:9} - \ref{fig:12} show} the estimated CATE versus the substitute of CATE (radius) of causal network (Causal Net), S-learner, T-learner, and adjusted regression with interaction term, for sample size $n= 10000$. The average substitute of CATE is around $8$. \hz{The left subplot of} Figure \ref{fig:7} shows the performance of different methods on noisy data when sample size varies from $2000$ to $10000$. The behavior on noisy image data is similar to that of noiseless image data: \hz{our causal network can detect heterogeneous treatment effect with high accuracy, while other methods cannot.}



\subsection{ Simple Relations  }

Since the linear regression based methods and tree based methods have been under intensive studied and has theoretical guarantees under simple model settings, we also want to test how the methods perform when the data generation procedure follows those simple models, or are likely to be preferable to linear regression base methods or tree based methods. We find four kinds of setting worth testing: data comes from a linear model, data comes from a polynomial model, data comes from two trees (one for treatment group and one for control group), data comes from two neural networks (one for treatment group and one for control group).

The covariates we use here are noisy images mentioned above, we generate outcomes according to different simple relations.

 \textit{Simulation Setting.} For all simple relations, data comes from two regression model and the experiment is randomized and balanced.  More precisely, it follows the following model.
\begin{itemize}
\item $Y_i(0)=f_0(X_i)+\epsilon_i(0)$,
\item $Y_i(1)=f_1(X_i)+\epsilon_i(1)$,
\item $\epsilon_i(0) \sim N(0,\sigma^2)$, $\epsilon_i(1) \sim N(0,\sigma^2)$,
\item $\epsilon_i(0)$, $\epsilon_i(1)$ and $X_i$ are mutually independent,
\item $X_i$ obeys the same distribution as of noisy image data in previous subsection \hz{and is vectorized,}
\item $T_i \sim B(1,0.5)$, which is independent of $(X_i, Y_i(0),Y_i(1))$.
\end{itemize}
\hz{We set the value of $\sigma$ such that the signal-to-noise ratio roughly equals 10, that is, 
$$
\sigma=\sqrt{ \frac{\sum_{i=1}^{n} \left( Y_i^\obs \right)^2}{10n}},
$$ 
where $Y_i^\obs,i=1,\cdots,n$ are the observed outcomes we generated in the training sample with sample size $n= 10000$.}

For linear data generator, 
$$
f_t(X_i)=X_i^T \beta_1 + t X_i^T \beta_2, \quad t = 0, 1,
$$
\hz{where $\beta_1 \in  \mathbb{R}^{1024}, \beta_2 \in \mathbb{R}^{1024}$ are regression coefficients whose first 20 components are generated from $N(0,10)$ (with seed 5) and the remaining ones are $0$.}



For polynomial data generator, 
$$
f_t(X)=X_i^T \beta_3 + X_i^T D_1X_i +  t ( X_i^T \beta_4 + X_i^T D_2 X_i) , \quad t = 0,1,
$$
where $D_1$ and $D_2$ are diagonal matrix. All components of $\beta_3,\beta_4, \textnormal{diag}(D_1), \textnormal{diag}(D_2)$ are generated independently from $N(5,50)$. 

For tree based data generator, $f_1$ and $f_2$ are random forests trained by mnist dataset, with responses being 1 to 9 and interchanging responses with same remainders when divided by 5. 


For neural network based data generator, $f_1$ and $f_2$ are pre-trained VGG net and pre-trained Alexnet respectively. 


 \textit{Results.} For all the Figures shown below, $y$ axis is MSE divided by the variance of CATE of test dataset. Therefore, the method is only \hz{useful} when it is less than 1, or it is meaningless to say the method could detect heterogeneity. When it is larger than 1, fine-grid comparison scale is not reasonable, going to scales as rough as the order it is of may be a better choice. 

\begin{figure}
\centering
\includegraphics[width=.8\textwidth]{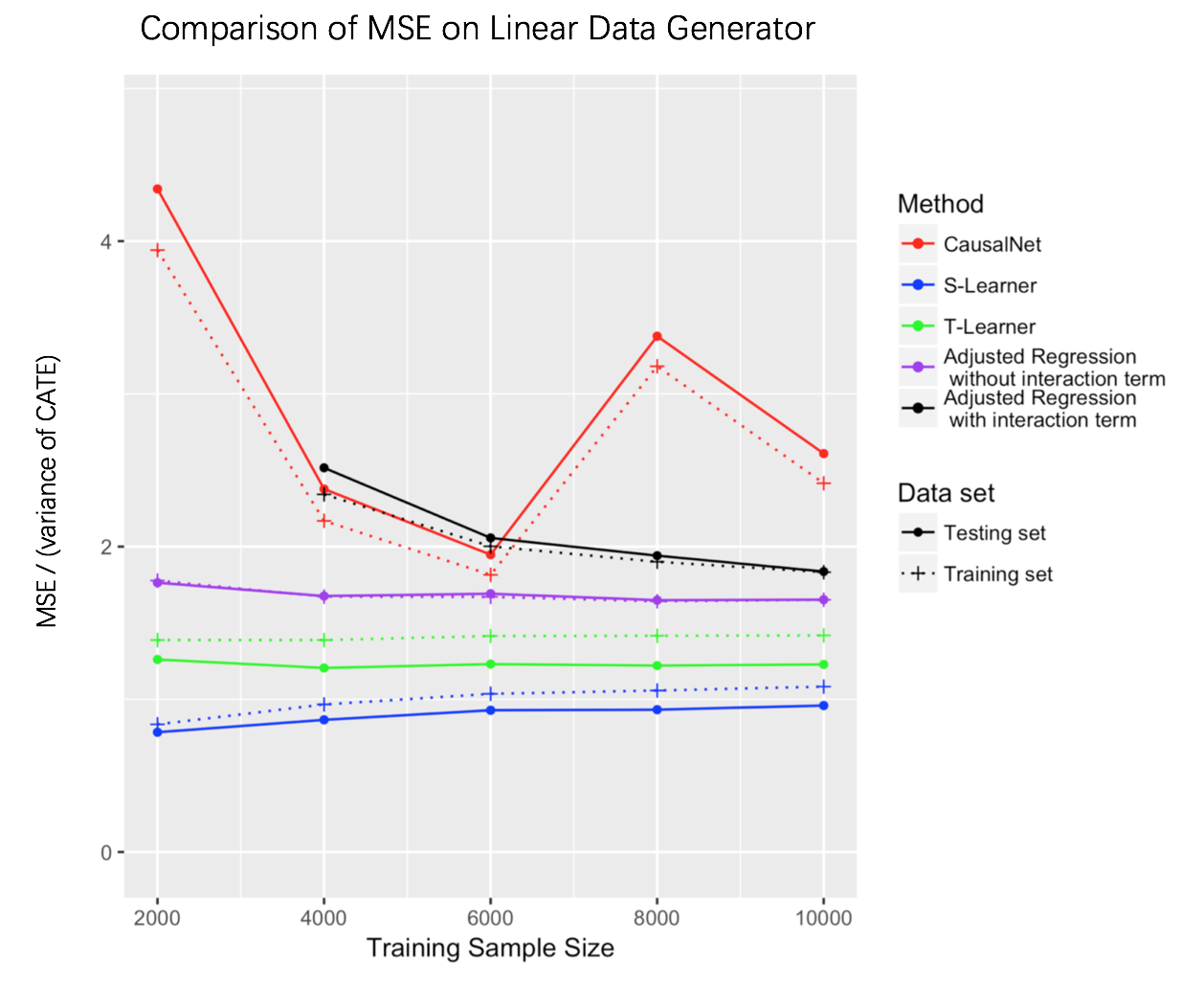}
\caption{Performance of five methods on linear data generator with five training sample size.}
\label{fig:14}
\end{figure}

\noindent\textit{Linear.} Figure \ref{fig:14} shows the performances of different methods, with different training sample size, on linearly generated data. Missing points stand for value being larger than the scope. In this setting, linear regression based methods could be theoretically proved to be optimal. However, tree based methods perform better here, which may due to the sparsity imposed during data generation. Causal network do not perform well in this setting, but it is of the same order with the best one, while the best one is no better than always guessing the average.

\begin{figure}
\centering
\includegraphics[width=.8\textwidth]{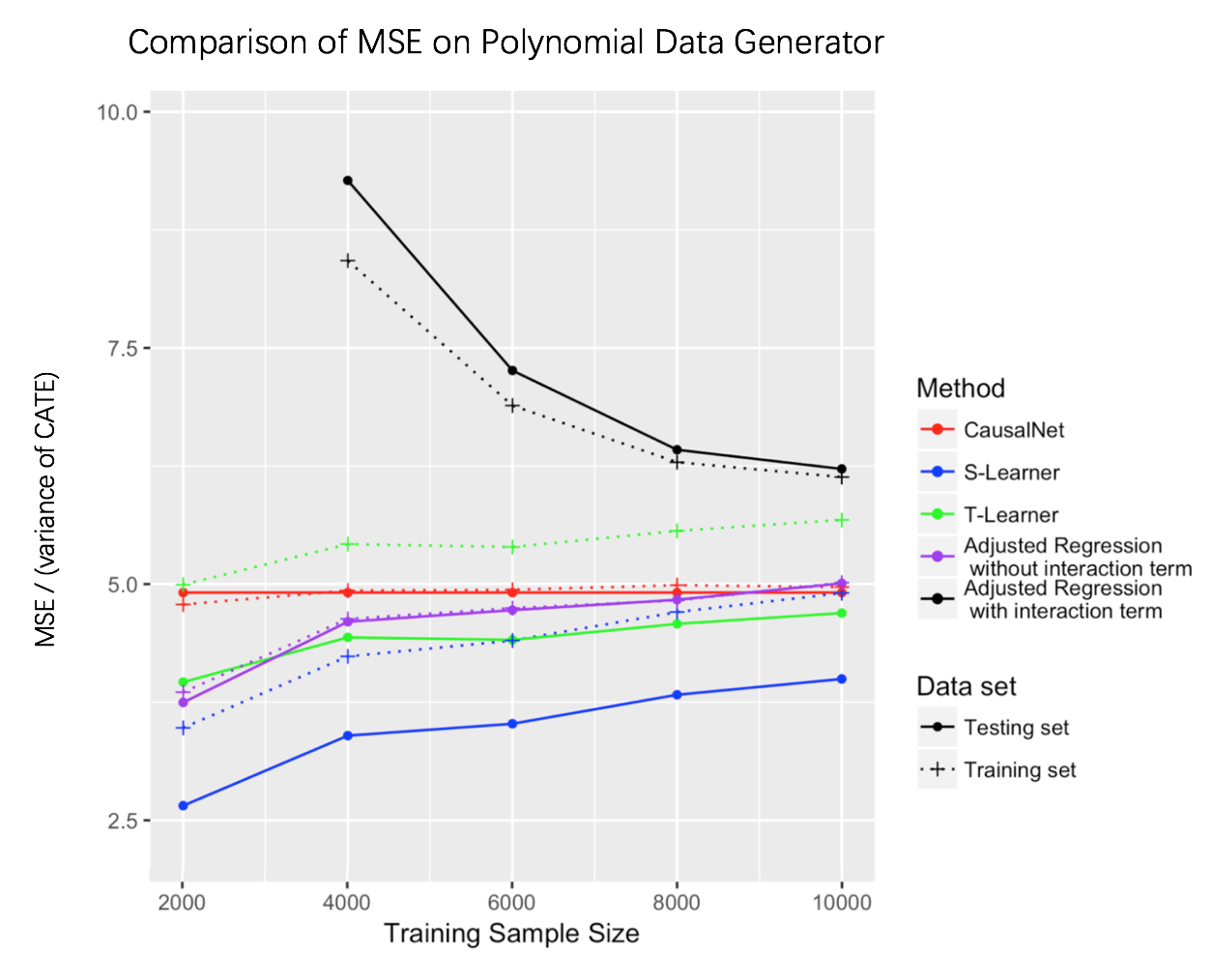}
\caption{Performance of five methods on polynomial data generator with five training sample size.}
\label{fig:15}
\end{figure}

\noindent\textit{Polynomial.} Figure \ref{fig:15} shows the performances of five methods, with different training sample size, on polynomially generated data. Missing points stand for value being larger than the scope. All the methods are well above one, indicating bad performances. They are also of the same order except adjusted regression with interaction term.

\begin{figure}
\centering
\includegraphics[width=.8\textwidth]{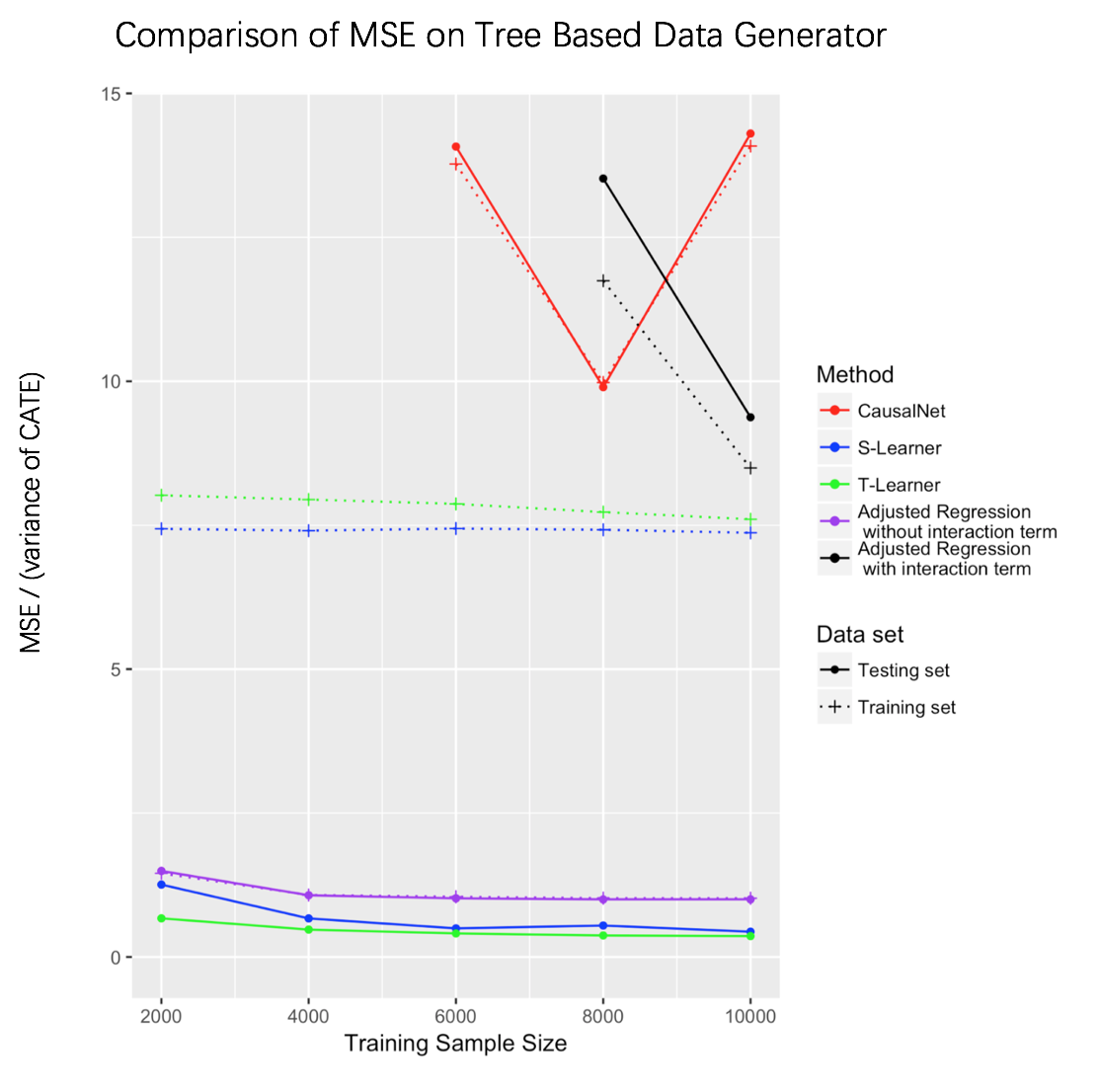}
\caption{Performance of five methods on tree based data generator with five training sample size.}
\label{fig:16}
\end{figure}

\noindent\textit{Tree Based Data Generator.} Figure \ref{fig:16} shows the performances of five methods, with different training sample size, on data generated by tree based data generator. Missing points stand for value being larger than the scope. This setting is expected to be favorable for trees, though tree based methods mysteriously have drastically higher training error than test error. Though both training error and test error of tree based methods are better than others, we can see that training error is of the same order of other methods and test error is also above 1. All the methods perform bad in this setting.

\begin{figure}
\centering
\includegraphics[width=.8\textwidth]{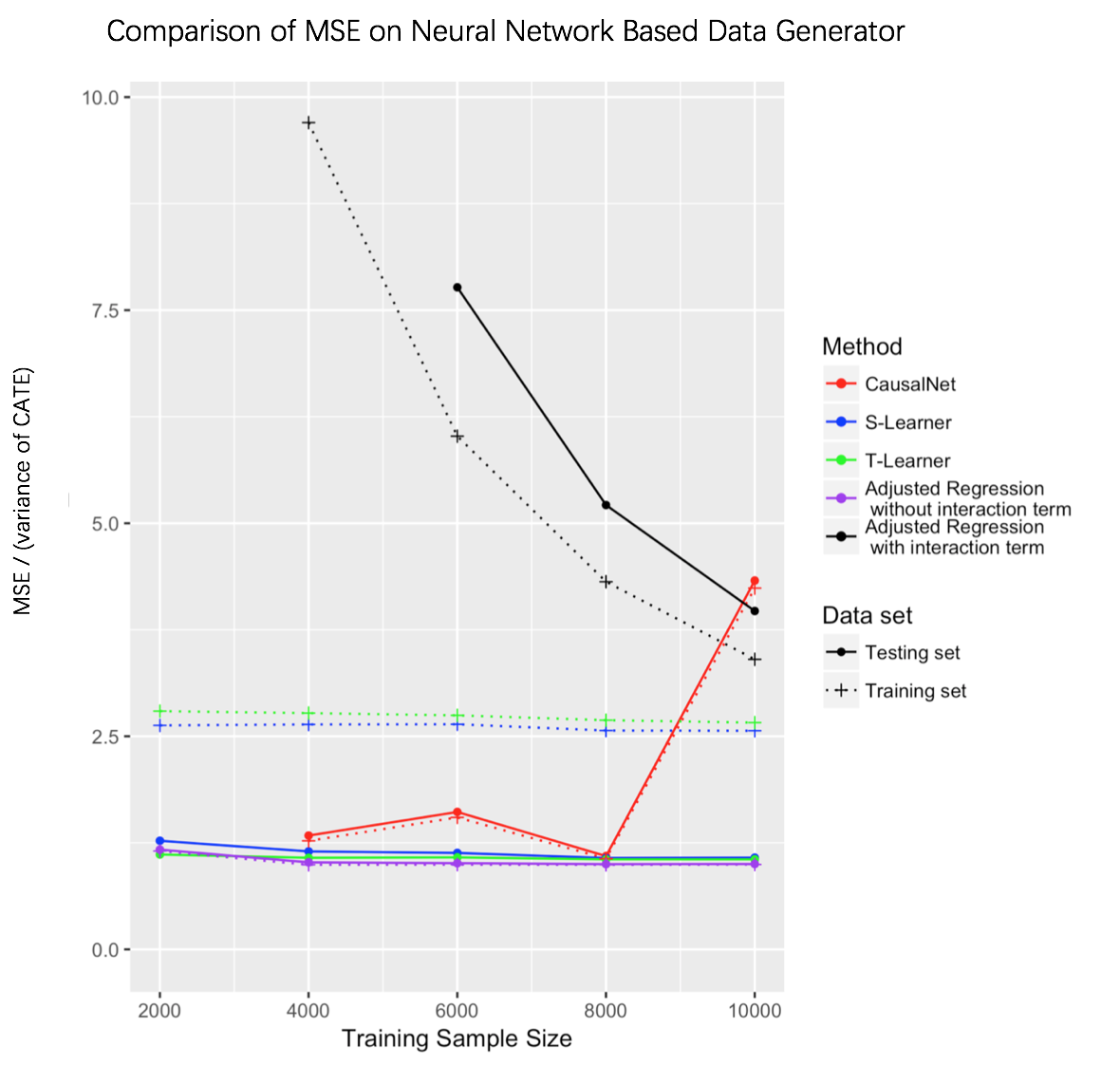}
\caption{Performance of five methods on neural network data generator with five training sample size.}
\label{fig:17}
\end{figure}

\noindent\textit{Neural Network Based Data Generator.} Figure \ref{fig:17} shows the performances of five methods, with different training sample size, on data generated by neural network based data generator. The missing points stand for value being larger than the scope. This setting is expected to favor neural network, however, since both VGG network and Alexnet are much more complicated and involved, causal network does not include those models, despite of the fact that all of them are neural networks. In this setting, all the methods performs unsatisfactory -- they are all above one, and of the same order, except the adjusted regression with interaction term.

\section{Discussion}

In this paper, we reinterpreted heterogeneous treatment effect estimation, discussed both the criterion for heterogeneous treatment effect estimation methods and the justifiability of using CATE to quantify heterogeneous treatment effect. 

We analyzed both the advantages and issues needing consideration of integrating neural network into heterogeneous treatment effect estimation. It wins in its expressiveness, ability of dealing with data of various forms (e.g. structured data like image and text), ease of combining different networks into one, short computational time for moderately large network, computational support on hardware level, and the ability for smart use of both shared and separate information of control and treatment group (our diverter mechanism). It losses in issues of its optimization, the accuracy and convergence behavior (optimization wise convergence) of which is not yet well understood and involves tuning (optimization wise), despite of the empirical success and wide applicable range of default off the shelf parameter setting.

We proposed diverter mechanism to automatically enable both information share and separation in treatment and control group, and in our diverter mechanism, treatment indicator is not by nature restricted to 0 and 1, it can take continuous value, like dose. We give a specific configuration of causal network. We tested our network on simulated image data, both in setting where the image (covariates) does not have completely irrelevant noise, and in setting where the image (covariate) has completely irrelevant noise, thus we can also see where CATE stops to be a good quantification of heterogeneous treatment effect. In both settings, causal network performs much better than other methods. We tested causal net along with linear regression based methods and tree based methods on simulated data generated by linear data generator, polynomial data generator, tree based data generator and neural network based data generator, with the covariate being the images mentioned above, in order to see how causal network behaves in settings where linear regression based method, tree based methods, neural network based methods, S-learner and T-learner are expected to do better. But the settings themselves are hard problem due to high dimensionality of images, the nonstandard distribution of covariate and topological-free relationship between the responses and images. We found that all the methods behave badly on these settings, and approximately equally bad except adjusted regression with interaction term being extremely bad, especially when training sample size is small.

Therefore, causal networks shows huge advantage in image data setting, where the topological structure is related to heterogeneous treatment effect. And for hard situations, our specific causal network does not save the day in all the cases, and does not ruins the day either.

Our specific causal network configuration is very simple and is CNN based, but it already shows huge advantage in heterogeneous treatment effect estimation with informative image data. It is promising to combine the strength of other time proved network structures (e.g., RNN) and expressiveness of networks (incorporating time proved non neural network based methods or structures) to fully borrow the strength of neural networks into heterogeneous treatment effect estimation with various form of data. Given the flexibility of integrating treatment information into the neural network, exploration of continuous treatment indicator or indicator of different forms is promising. In the end, however, it is always important to balance the representation capacity, the network structure complexity and computational complexity, and this should always be kept in mind when designing networks.

\section{ Acknowledgment}
Part of the work is part of the first author's undergrad thesis at Tsinghua University, the first author wants to thank Professor Michael Zhu, for his machine learning seminar held in Tsinghua University; Karl Kumbier and Professor Bin Yu, for introducing the first author to state of art machine learning methods during her visit to Berkeley. The first author also wants to thank the stat department of Wharton School, for supporting her PhD study. \hz{Dr. Hanzhong Liu's research is supported by the National Natural Science Foundation of China grant 11701316.}

\section{Appendix}

Following are some exploratory simulations to explore neural networks' characteristics on simplified toy networks. 
\subsection{General Experiment Setting}
\subsubsection{Data Generator}
We have two data generation settings basically representing different level of nonlinearity --- linear data generator, and polynomial data generator. Changing the variance parameter in polynomial data generator can give different level of nonlinearity, with growing nonlinearity as variance grows. Details are shown as follows.
\begin{itemize}
\item Linear Data Generator\\
	The covariate vector ($X$) is of dimension 9. The data generating mechanism for $i$-th sample is:
	$$Y_i = \sum_{j=1}^9 jX_i^{(j)} + 10 T_i + \epsilon_i .$$
	Where $X_i^{(j)} \overset{i.i.d.}{\sim} N(0,\sigma^2)$ for $1 \le j\le5 $, $X_i^{(j)} \overset{i.i.d}{\sim} U(0,5)$ for $6 \le j\le 9 $, $\epsilon_i \overset{i.i.d}{\sim} N(0,1)$, and the treatment indicator $T_i \overset{i.i.d}{\sim} B(1,0.5)$. Each covariate is generated independently from other covariates and the treatment indicator.
\item Polynomial Data Generator\\
	The covariant ($X$) is of dimension 9. The data generating mechanism for $i$-th sample is:
	$$ Y_i= \sum_{j=1}^9 (jX_i^{(j)}+(11-j){X_i^{(j)}}^2) + 10 T_i +\epsilon_i . $$
	Where $X_i^{(j)} \overset{i.i.d}{\sim}N(0,\sigma^2) $ for $1 \le j\le5 $, $X_i^{(j)} \overset{i.i.d}{\sim} U(0,5)$ for $6 \le j\le 9 $, $\epsilon_i \overset{i.i.d}{\sim} N(0,1)$, and $T_i \overset{i.i.d}{\sim} B(1,0.5)$. Each covariate is generated independently from other covariates and the treatment indicator.
\end{itemize}


\subsubsection{Networks}
Networks included in our exploration are:
\begin{itemize}
\item Linear Neural Network\\
	One layer neural network with only linear operator, the form of which is the same as linear regression.
\item Linear with Non Linear Activation Function \\
	One layer neural network with linear operator and sigmoid activation function.
\item Two Layer Nonlinear Activation Function \\
	Two layer neural network with sigmoid activation function following a batch normalization between two layers.
\end{itemize}

\subsubsection{Evaluating Method}
Since we use Adam to be our optimizer for neural network, the parameters of the neural network is updated after each mini-batch optimization iteration, we consider how the neural network estimator performs with increasing epoch (the iteration number of Adam). One thing to notice is that, though for each epoch we generate an independent mini batch with our data generator, which seems to require a very big sample size, we do not necessary need the sample size to be the same number as the data used in optimization procedure, as each sample can be used many times in the iteration procedure --- we can keep subsampling from the original sample to continue the optimization procedure. We keep the mini batch to be 200, and the sample size for adjusted regression also be 200. This is not a fair game as mentioned before (sample size are not the same), but we look at the trend, which does not change under scaling.

\subsection{Linear Neural Network on Linear Data generator}
\begin{figure}[!h]
\centering
\includegraphics[width=0.6\textwidth]{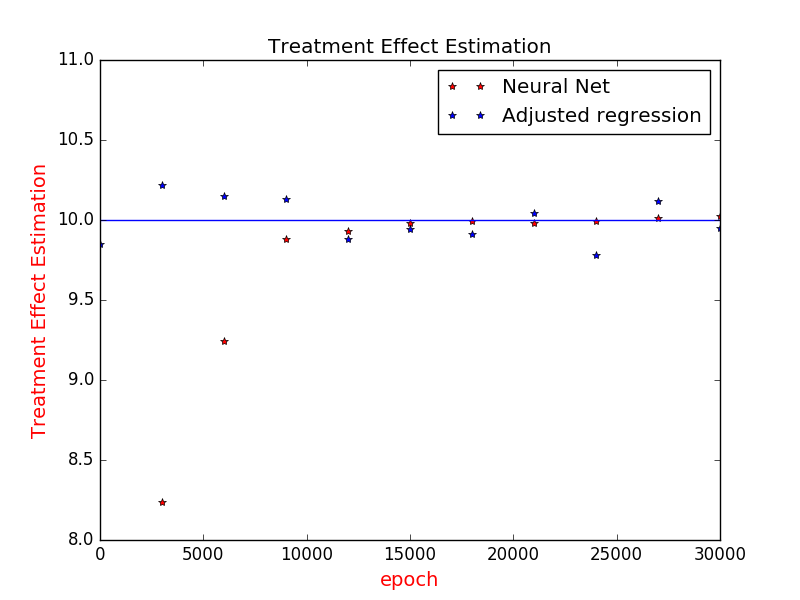}
\caption{Computing accuracy}
\label{fig:18}
\end{figure}

\begin{itemize}
\item Network: One layer linear neural network with linear activation function
\item Data Generator: Linear Data Generator (with $\sigma=10$)
\item Purpose: Separate out computational issue
\item Result: Figure \ref{fig:18} shows the computational issue affects accuracy, but when sample size gets larger and iteration number gets larger, it is alleviated. 
\end{itemize}

\subsection{Linear with Non Linear Activation Function}
\begin{figure}[!h]
\includegraphics[width=0.45\textwidth]{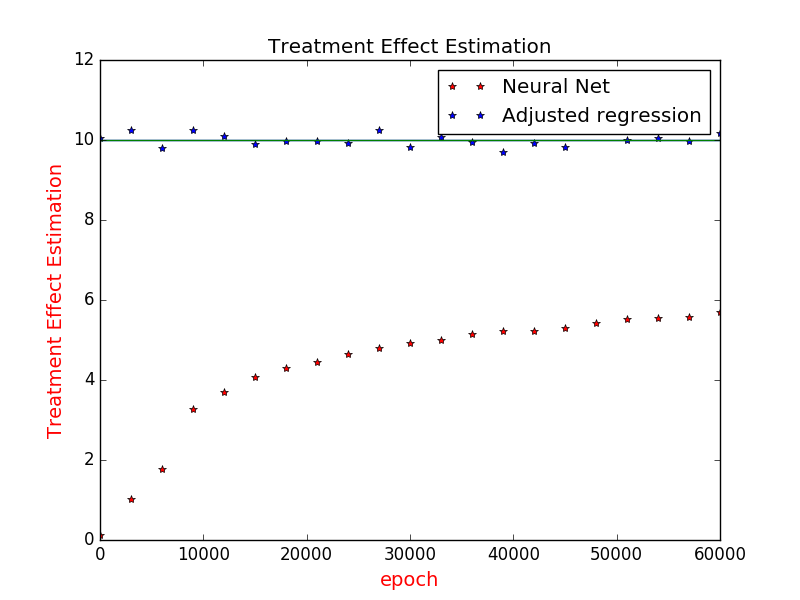}
\includegraphics[width=0.45\textwidth]{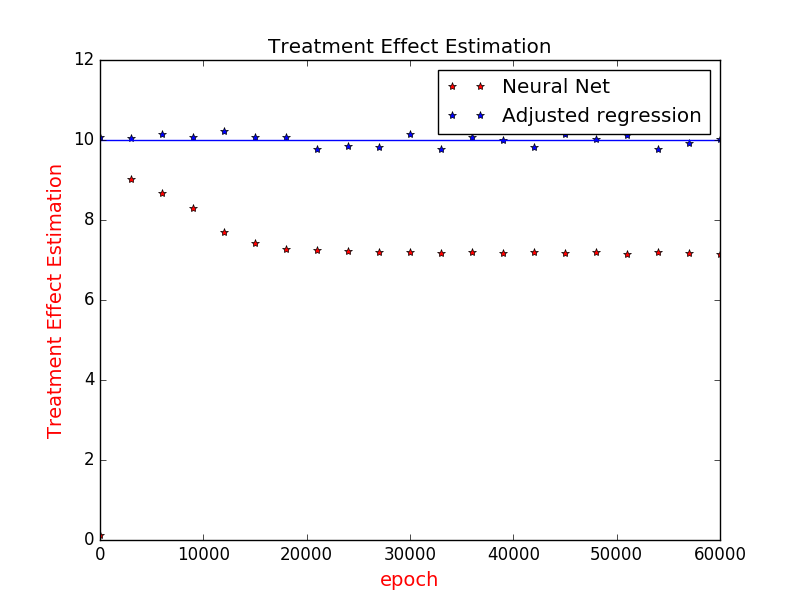}
\caption{Influence of starting points}
\label{fig:19}
\end{figure}

\begin{itemize}
\item Network: One layer linear neural network with sigmoid activation function
\item Data Generator: Linear Data Generator (with $\sigma=1$)
\item Two optimizing initializing points
\item Purpose: Separate out issue of choosing initializing points
\item Result: Figure \ref{fig:19} shows different initial points affects the accuracy in the first several iterations but in the long run, is not a big issue.
\end{itemize}

\subsection{Ability of detecting nonlinearity}
\begin{figure}
\centering
\includegraphics[width=0.6\textwidth]{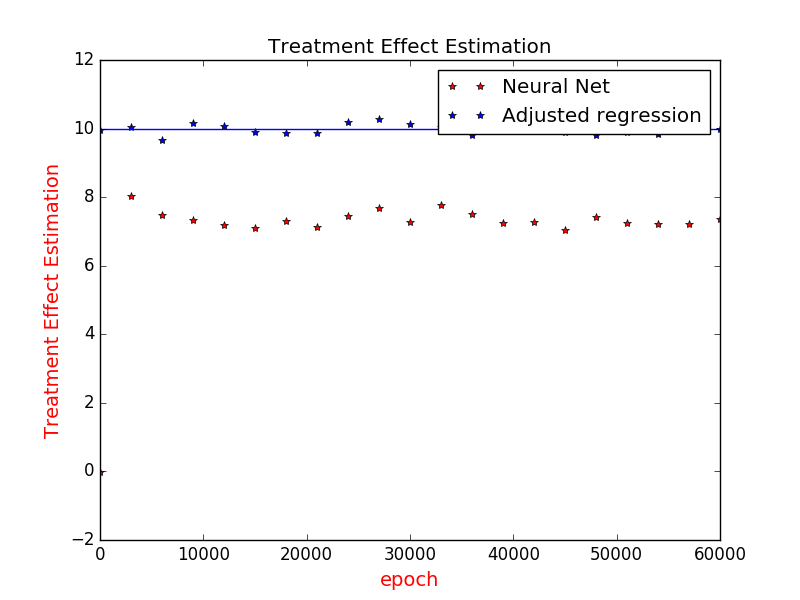}
\caption{Two layer nonlinear NN with Linear Data generator}
\label{fig:21}
\end{figure}

\begin{figure}
\centering
\includegraphics[width=0.6\textwidth]{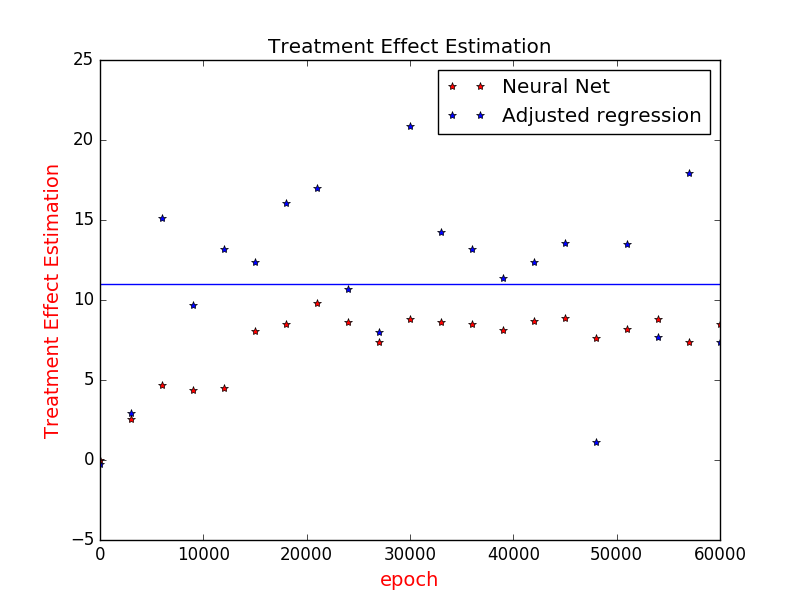}
\caption{Two layer nonlinear NN with Nonlinear Data generator, Small Variance}
\label{fig:22}
\end{figure}

\begin{figure}
\centering
\includegraphics[width=0.6\textwidth]{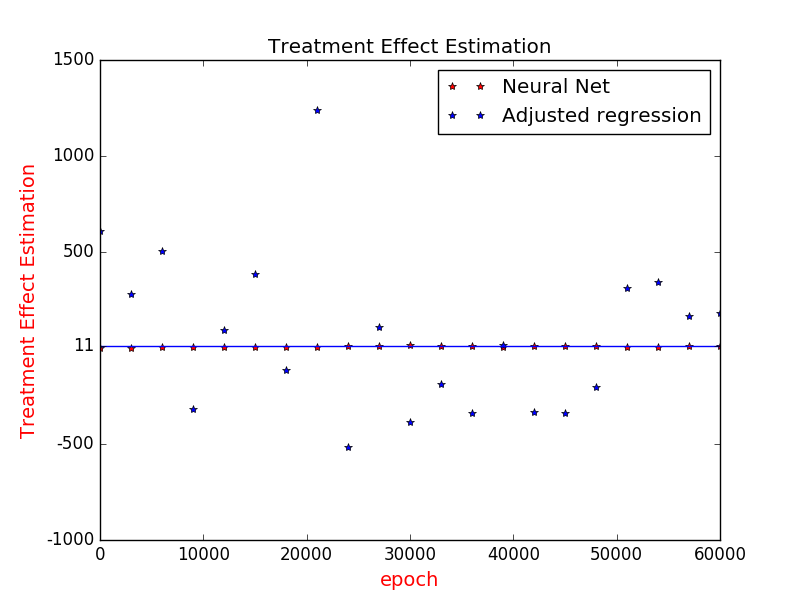}
\caption{Two layer nonlinear NN with Nonlinear Data Generator, Large Variance}
\label{fig:23}
\end{figure}

\begin{itemize}
\item Network: Two layer nonlinear activation function
\item Data Generators: Linear (Figure \ref{fig:21}, $\sigma=1$), polynomial with small variance (Figure \ref{fig:22}, $\sigma=1$), polynomial with large variance (Figure \ref{fig:23}, $\sigma=10$)
\item Purpose: Explore the behavior of deeper neural network on linear data; Explore the ability of detecting nonlinearity
\item Results: Two layer nonlinear activation function can detect nonlinearity when linearity increase, though perform a bit worse than linear regression based method in linear model, in which setting the linear regression based method is provably optimal.
\end{itemize}

\end{document}